\documentclass[pdflatex,sn-nature]{sn-jnl}
\usepackage{mathrsfs}

\usepackage{mathrsfs}%
\usepackage{graphicx}%
\usepackage{multirow}%
\usepackage{amsmath,amssymb,amsfonts}%
\usepackage{amsthm}%
\usepackage{mathrsfs}%
\usepackage[title]{appendix}%
\usepackage{xcolor}%
\usepackage{textcomp}%
\usepackage{manyfoot}%
\usepackage{booktabs}%
\usepackage{algorithm}%
\usepackage{algorithmicx}%
\usepackage{algpseudocode}%
\usepackage{listings}%
\usepackage{outlines}%

\theoremstyle{thmstyleone}%
\theoremstyle{thmstyletwo}%

\theoremstyle{thmstylethree}%

\begin{document}

\title{Expanding Protein Structure Prediction into Conformational State Space}

\author[1]{Devlina Chakravarty}
\author[2]{Justin J. Miller}
\author[3]{Da Teng}
\author[4,5]{Yousuf O. Ramahi}
\author[6]{Patrick Bryant}
\author[7,8]{Camila Neira-Mahuzier}
\author[7,8]{César A. Ramírez-Sarmiento}
\author[4,5,9]{Sarah Rauscher}
\author[2]{Gregory R. Bowman}
\author[10,3,11]{Pratyush Tiwary}
\author*[1,12]{Lauren L. Porter\email{porterll@nih.gov}}

\affil[1]{Division of Intramural Research, National Library of Medicine, National Institutes of Health}
\affil[2]{Department of Biochemistry and Biophysics, University of Pennsylvania}
\affil[3]{Institute for Health Computing, University of Maryland}
\affil[4]{Department of Chemical and Physical Sciences, University of Toronto, Mississauga}
\affil[5]{Department of Chemistry, University of Toronto}
\affil[6]{Department of Molecular Biosciences, Stockholm University}
\affil[7]{Institute for Biological and Medical Engineering, Schools of Engineering, Medicine and Biological Sciences, Pontificia Universidad Católica de Chile}
\affil[8]{ANID, Millennium Science Initiative Program, Millennium Institute for Integrative Biology (iBio)}
\affil[9]{Department of Physics, University of Toronto}
\affil[10]{Department of Chemistry and Biochemistry, University of Maryland, College Park}
\affil[11]{Institute for Physical Science and Technology, University of Maryland, College Park}
\affil[12]{National Heart, Lung, and Blood Institute, National Institutes of Health}
\date{}

\renewcommand{\thesection}{\Roman{section}}
\renewcommand{\thesubsection}{\arabic{section}.\arabic{subsection}}

\maketitle

\begin{abstract}

\textbf{Abstract\newline}
   Recent AI advances have enabled protein structure prediction at near-experimental accuracy, largely solving the problem of identifying a dominant conformation from sequence. Many proteins, however, function as dynamic systems populating multiple conformational states with activity emerging from shifts in relative occupancy—an incomplete picture when reduced to one structure. Here, we argue that structure prediction should be reformulated as a state-space inference problem: recovering not one conformation's coordinates but accessible states, their energetic and kinetic relationships, context dependence, and responses to perturbations. We review emerging strategies—deep learning ensemble generators, physics-based simulations, and experimental constraints—and outline a roadmap toward state-space prediction.
\end{abstract}

\section{The Success of the Single-Structure Paradigm and Why It Is No Longer Sufficient}

Protein structure prediction has reached a historic milestone. Anfinsen's thermodynamic hypothesis \cite{RN1} — that sequence encodes structure — has culminated in deep learning (DL) systems that predict a representative structure with near-experimental accuracy \cite{RN38}. This success has been transformative: predicted structures now suggest protein function, rationalize mutations, initialize drug design and protein engineering campaigns, and support experimental structure determination and molecular dynamics (MD) studies.

\begin{figure}
    \centering
    \includegraphics[width=1.0\linewidth]{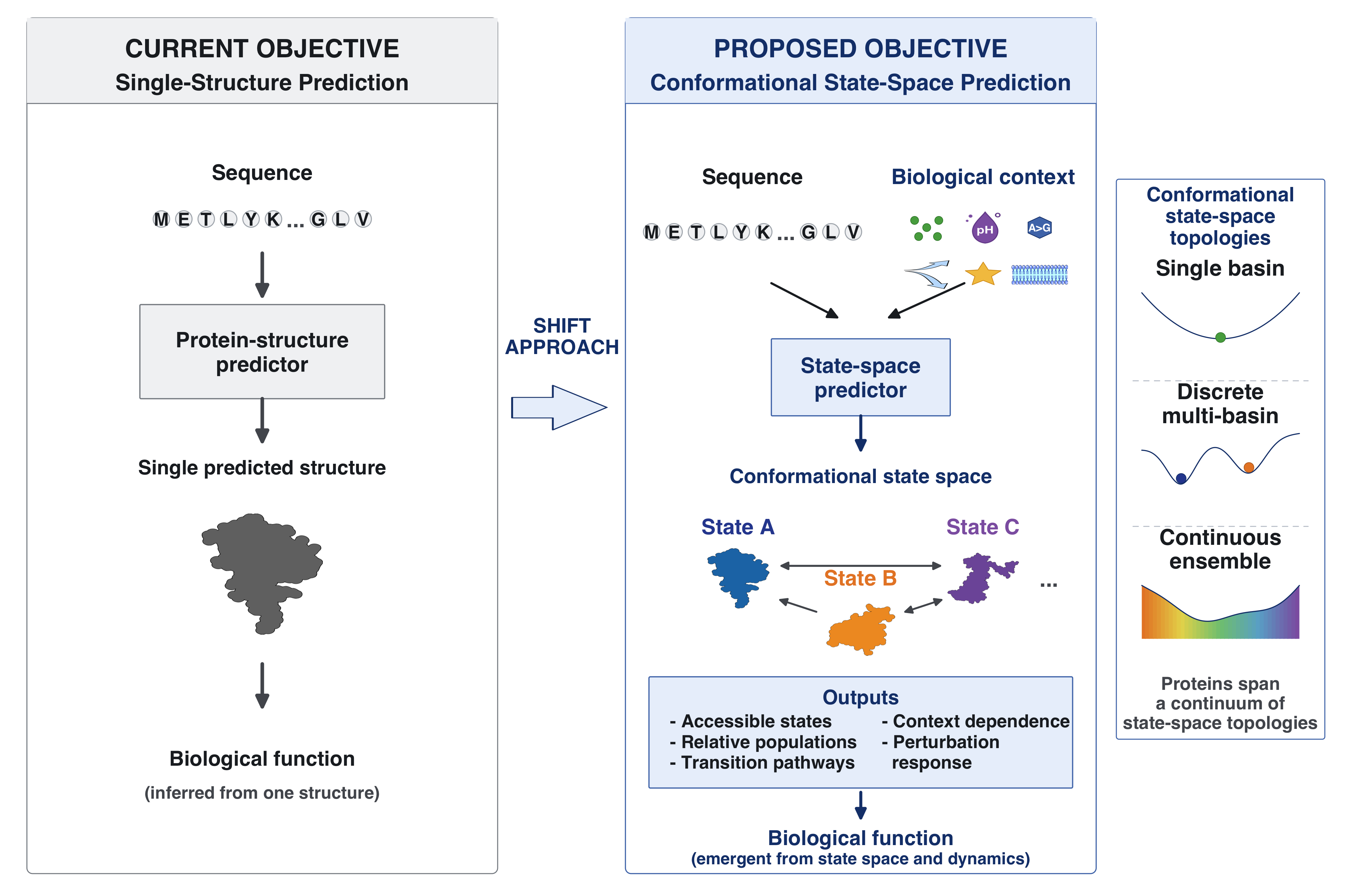}
    \caption{\textbf{\textbf{Shifting the objective of protein structure prediction from single structures to conformational state spaces.}} \textit{(Left)} The current paradigm predicts a single structure from sequence alone, from which biological function is inferred. \textit{(Right)} The proposed paradigm conditions prediction on both sequence and biological context (e.g., ligands, pH, post-translational modifications, binding partners, membrane environment) to output a full conformational state space — the accessible states (A, B, C), their relative populations, the pathways connecting them, their dependence on context, and their response to perturbation. Biological function emerges from this state space and its dynamics, rather than being inferred from a single coordinate model. \textit{(Right inset)} Proteins are not uniformly multi-state in the same way: state spaces range from a single dominant basin, to a small number of discrete alternative basins, to a continuous ensemble of interconverting conformations, and a general framework must accommodate this full range of topologies.
}
    \label{fig:fig1}
\end{figure}

Yet the computational problem solved by DL is narrower than the biology it seeks to explain. Current methods predict a single low-energy structure, with maximum coordinate accuracy, whereas most proteins populate ensembles of interconverting conformations spanning simple single-basin systems to complex multi-state and intrinsically disordered landscapes (Figure \ref{fig:fig1}). A single structure suffices for the simple but not the complex.

These alternative states are frequently not noise but functionally essential: they govern domain motion, transport, substrate recognition, and regulatory switching \cite{RN4, RN7, RN115, RN107}, and many biological functions depend specifically on low-population, higher-energy states \cite{RN89, RN113, RN114}. Although MD simulations can probe these states, they remain computationally expensive and depend on accurate starting structures \cite{RN120, RN122}. 

Current training data reinforce this limitation. Although the Protein Data Bank (PDB) contains more than 250,000 structures, it is heavily biased toward stabilized conformations, with relatively few experimentally characterized alternative states \cite{RN26,RN25}.  X-ray crystallography, cryo-electron microscopy (cryo-EM), and solution nuclear magnetic resonance (NMR) spectroscopy each preferentially observe subsets of the underlying conformational landscape, while ensemble-sensitive approaches such as hydrogen-deuterium exchange-mass spectrometry (HDX-MS), small angle X-ray scattering (SAXS), and single-molecule methods increasingly reveal heterogeneity invisible to single coordinate models \cite{RN26,RN25,RN124,RN125,RN27,RN28,anthis2015visualizing,RN30,RN32,RN31}.

The central question is thus no longer `\textit{What is the structure?}'' but ``\textit{What structural ensemble of states does this sequence encode, under what conditions, with what populations, and how do they interconvert?}'' (Figure \ref{fig:fig1}) Single-structure prediction is therefore a special case of the broader landscape-inference problem. Biophysicists have long approached folded proteins this way, and recognized a continuum from extremely rigid proteins to intrinsically disordered proteins (IDPs) \cite{RN111}, all of which must be understood through \textbf{conformational state spaces}. Unlike ensemble prediction, which estimates conformational populations under a specific condition, state space prediction seeks to infer the organization of accessible states, their energetic relationships, the transitions connecting them, and how those populations are reweighted by mutations and changes in biological context.  The next breakthrough will come not from improving coordinate accuracy of a single conformation, but from predicting how sequences encode the dynamic organization of accessible states across biological contexts.

In this perspective, we propose that the objective of protein structure prediction must shift from recovering the coordinates of a single conformation to inferring a conformational state space. We first identify the biological and physical boundary conditions under which the single-structure paradigm breaks down (Section II), discuss the importance of modeling conformational state spaces (Section III), and review current approaches that recover complementary components of conformational state space (Section IV). Finally, we outline a roadmap toward state-space prediction (Section V), arguing that future predictors must integrate alternative-state generation, energetic modeling, and biological context into a unified inference framework.

\section{Limitations of single-structure predictions}

\begin{figure}
    \centering
    \includegraphics[width=1.0\linewidth]{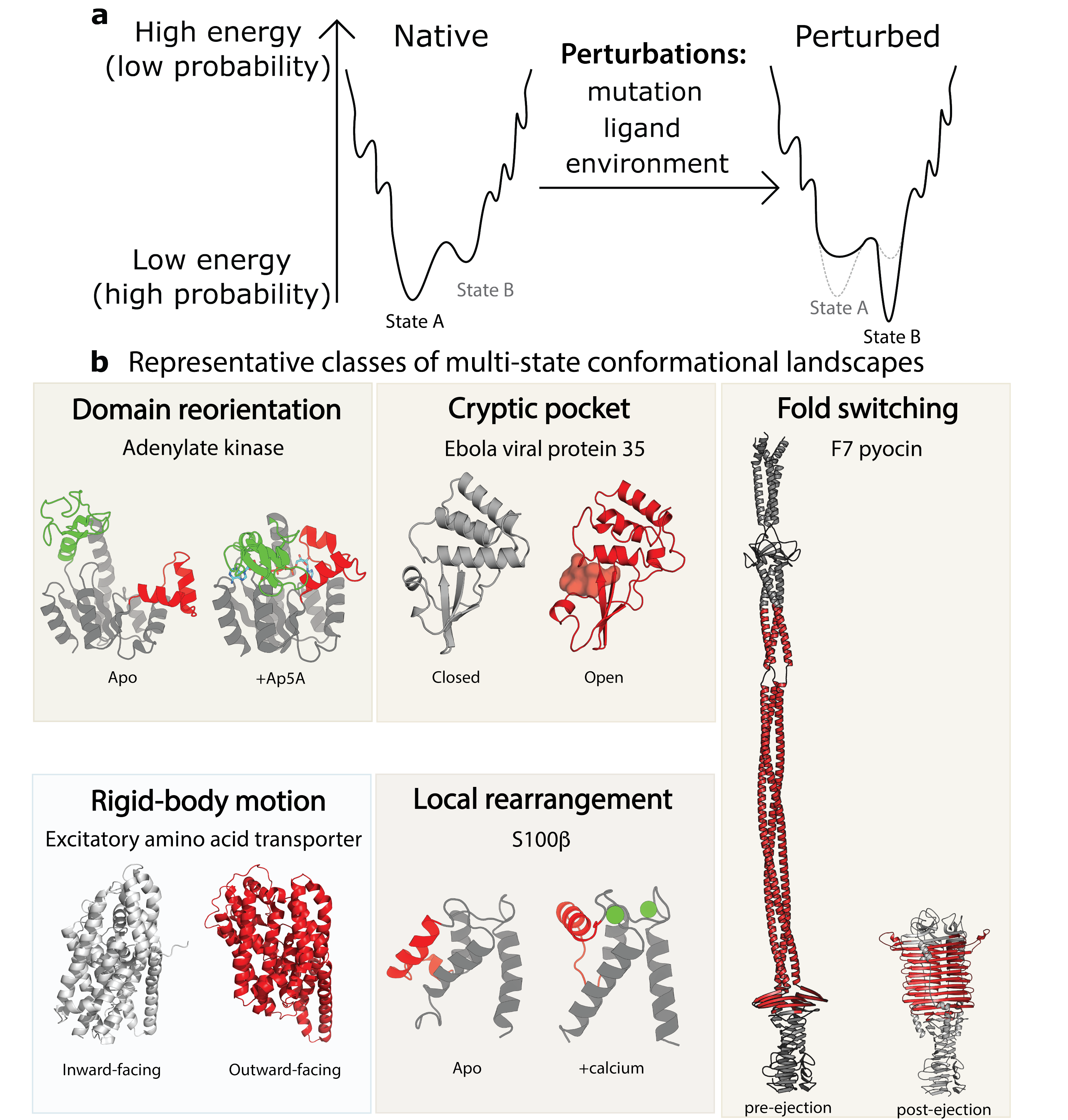}
    \caption{\textbf{Representative classes of multi-state conformational landscapes.} \textit{(a)} Environmental perturbations—including mutations, ligand binding, and changes in physicochemical conditions—reshape the underlying conformational landscape by altering the relative stability of accessible states. In this example, a perturbation shifts the equilibrium from a dominant State A to an alternative State B, illustrating that biological function often depends on changes in state populations rather than the emergence of entirely new conformations. \textit{(b)} Representative biological manifestations of multi-state conformational landscapes. Examples include domain reorientation (adenylate kinase), rigid-body motion (excitatory amino acid transporter), cryptic-pocket formation (Ebola virus protein 35), local structural rearrangement ($S100\beta$), and topological fold switching (F7 pyocin). Together, these examples illustrate distinct boundary conditions under which a single coordinate model becomes an incomplete representation of the biologically relevant conformational state space.}
    \label{fig:fig2}
\end{figure}

The limitations of single-structure prediction emerge under specific conditions defined by the conformational landscape. The central issue is not whether a dominant conformation exists, but whether a single set of coordinates can adequately capture the biologically relevant state space. In simple systems, where the free-energy landscape is dominated by a single basin, a representative structure is often a sufficient approximation. However, this assumption breaks down when multiple basins, kinetic barriers, or context-dependent effects contribute to function. The modern view of protein folding emphasizes ensembles of states rather than a single endpoint \cite{RN9,RN10,RN11}. Although the energy landscape is often shown as a funnel that guides proteins toward a folded structure, many proteins differ from this simple model. In these cases, the conformational landscape becomes multi-dimensional, and protein function depends on how the states are spread across it (Figure \ref{fig:fig2}).

\subsection{Multi-Basin Landscapes with Topological Continuity}
A first boundary condition is when multiple metastable basins exist with the overall protein fold remaining similar. These include local conformational rearrangements, such as side-chain reorientation, loop remodeling, and small secondary-structure fluctuations, as well as larger rigid-body domain motions. Although these transitions preserve the global topology, they can yield discrete functional states with distinct geometric and energetic properties. 

For example, adenylate kinase (ADK) has three domains, two of which close against a core domain upon substrate binding.  This closing motion is the rate-limiting step for ADK's catalytic activity \cite{RN103}.  Similarly, substrate binding in the excitatory amino acid transporter (EAAT) induces a $\sim$15~\AA{} rigid-body displacement of the transport domain relative to a stable scaffold domain, generating outward- and inward-facing states. TM-score comparisons indicate substantial motion in the transport domain (TM-score = 0.608), while the scaffold (0.834) and core structure (0.897) remain relatively stable \cite{RN13,RN12}. These transitions act as functional switches linking substrate transport to ion channel regulation (Figure \ref{fig:fig2}). 
The \textmu{}-opioid receptor exhibits ligand-dependent changes in the populations of side-chain rotameric states, priming the protein for either G protein or $\beta$-arrestin binding, as shown by nuclear magnetic resonance (NMR) spectroscopy and MD simulations \cite{RN116}. 
In such systems, a single structure models only one metastable basin, ignoring alternative states that can be important under physiological conditions. Even more subtly, minute allosteric effects can dramatically change function without large structural changes \cite{RN117}. In these cases, ensemble properties--not static ones--drive regulation.

\subsection{Rare or Low-Occupancy Excited States}
A second boundary condition arises when the function depends on low-population conformations. These states often correspond to shallow basins or excited sub-states that are transient and poorly captured in experimental structures. Cryptic pockets serve as a clear example. These transient binding sites are not visible in ground-state structures but are accessible via ensemble-sensitive measurements such as NMR and MD simulations, and in some cases crystallography \cite{zhang_decrypting_2026}. Ligand binding can stabilize cryptic pockets, change protein activity and regulate distal portions of the protein\cite{RN15,RN16,RN89,RN91}.  For example, the interferon inhibitory domain of the Ebola viral protein 35 harbors a cryptic pocket that allosterically controls the probability and mode of RNA binding \cite{RN15,RN90} (Figure \ref{fig:fig2}). Cryptic pocket opening has also been linked to enzyme activity \cite{RN89}, used as the basis for the first FDA approved KRAS inhibitor \cite{RN91,RN92}, and is a potential means of preferentially drugging protein isoforms with highly similar active sites \cite{RN93}. Two challenges are posed by these systems, first predicting the less populated conformational states where these pockets are present, as well as predicting the occupancy of these conformational states. A single high-confidence ground-state model is therefore insufficient to capture the mechanism.

\subsection{Topologically Distinct Basins: Fold Switching}
The strongest boundary condition arises when different basins correspond to topologically distinct folds. In fold-switching or metamorphic proteins, the same or nearly identical sequence can adopt multiple stable folds with different secondary structures, demonstrating that the sequence-single structure mapping is inherently flawed. These systems show that alternative basins can be evolutionarily encoded \cite{schafer2023evolutionary}, kinetically controlled, and functionally integrated. In such cases, the single-structure objective fails not because of prediction error, but because the target itself requires at minimum two divergent answers. Importantly, these fold-switching proteins are not biomolecular oddities, but rather abundant and widespread \cite{RN85}.\newline
A striking example of fold switching is the F7 pyocin, a phage-tail-like bactericidal nanomachine produced by \textit{Pseudomonas aeruginosa} \cite{RN75}.  Upon ejection from the nanomachine, a 163-residue segment of its central tail fiber undergoes a dramatic transition from a trimeric $\alpha$-helical coiled-coil to a triangular $\beta$-prism (Figure \ref{fig:fig2}). This fold switch is essential for bactericidal activity, remodeling the tail tip, ejecting the tape measure protein, and driving membrane puncture. Mutations that destabilize the $\beta$-prism abolish activity without affecting particle assembly, directly linking the conformational change to function. 

\subsection{Ion binding and pH}

Beyond ligand binding, ions are key environmental regulators of protein conformation, yet their identity, occupancy, and pH are absent from sequence-only structure-prediction inputs. Rather than adopting a single structure, many proteins shift among conformational basins in response to ionic conditions. For example, calcium binding reorganizes EF-hand coordination geometry and reorients neighboring helices, exposing hydrophobic surfaces for target recognition in S100 proteins \cite{Trave1995, Smith1998, Wright2005} (Figure \ref{fig:fig2}). Protons likewise modulate protein structure by altering the protonation states of titratable residues \cite{Swanson2007}. In the influenza A M2 proton channel, protonation of the His37 tetrad at low pH stabilizes more open, hydrated pore conformations that enable proton conduction \cite{Liang2016}. Proton-coupled transporters such as EmrE provide an even stronger example: protonation of active-site residues drives alternating conformational states that are essential intermediates of the transport cycle rather than structures favored under different conditions \cite{Morrison2015}. Together, these examples illustrate that environmental factors can both reshape conformational equilibria and drive obligatory transitions among multiple functional states, highlighting a fundamental limitation of sequence-only, single-structure predictions.

\subsection{Kinetic Structure and Context Dependence}
Across these regimes, thermodynamic stability alone does not determine biological relevance. Barrier heights, switching timescales, and environmental perturbations govern the accessibility and occupancy of states. For example, proline isomerization can occur on timescales of minutes to hours, creating kinetic bottlenecks that regulate  function. In the anti-HIV antibody 10E8, slow cis-trans isomerization reduces the population of the active conformation, thereby limiting binding efficiency \cite{RN20}. Fold-switching transitions can span from milliseconds to hours \cite{RN21,RN22,RN24,RN23}.  These systems demonstrate that kinetics, rather than thermodynamics alone, can encode biological function. Environmental conditions such as ligand binding, temperature, and macromolecular interactions can shift basin depths and change equilibrium distributions \cite{feng2019quantifying, miklos2011protein}. Mutations can likewise remodel landscapes by stabilizing alternative basins or modifying the barriers between them. For example, experimental data have demonstrated that targeted mutations that disrupt the stability of the $\beta$-prism in F7 pyocin abolish its activity, thereby revealing a direct connection between the fold-switching process and its function. Complicating matters further, codon frequency can dramatically shift a protein's fold and function \cite{sander2014expanding}.  Taken together, when a function relies on conditional occupancy or switching dynamics, the relevant object is not a static structure but a context-dependent distribution over states and the transitions connecting them. Predictive models should also have dynamics, barrier heights and timescales of functionally relevant motions as their targets, not only static structures or multi-state predictions. This type of information is already accessible to MD simulations, and MD datasets present a valuable source of training data for such models.

\section{Biological payoff for modeling conformational state spaces accurately}

Expanding protein structure prediction from single conformations to conformational state spaces is not merely a technical refinement but a change in the kinds of biological questions computational models can answer. Rather than asking only what structure a protein adopts, state-space prediction asks how proteins encode mechanism through alternative states, their populations, and their responses to perturbation. A single coordinate model identifies one likely structure; a conformational state space captures how proteins encode regulation, specificity, and adaptability through alternative states, their populations, and their responses to perturbation. Incorporating biological context—including environmental conditions, experimental restraints, and intrinsic sequence variation—allows prediction to move beyond describing structures toward explaining mechanisms. This expanded objective enables several classes of biological inference that remain inaccessible to single-structure predictors.
\newline

\noindent\textbf{\textit{Distinguishing alternative conformations and functions among homologous proteins.}} For decades, homologous proteins were assumed to adopt similar structures and functions \cite{RN56}. However, an increasing body of structural data has revealed that homologs can evolve into distinct folds with different functions \cite{RN57,RN58}. Because current prediction models rely heavily on multiple sequence alignments (MSAs), similar sequences frequently produce similar structural priors even when they experimentally occupy different conformational basins \cite{RN59}. This limitation is evident in several systems. In the human oncoprotein BCCIP, two functionally distinct isoforms sharing approximately 80\% sequence identity adopt structures differing by more than 10 Å, yet AlphaFold2 and AlphaFold3 predict nearly identical models because of training-set bias \cite{RN37,RN60}. Similar behavior is observed for pro-interleukin-18, DZZB, and MP20, where models converge on known folds despite experimental evidence for alternative conformations \cite{RN49,RN58,RN61,RN62}. These examples illustrate that homologous sequences may encode distinct conformational state spaces rather than simply different coordinate models.
\newline

\textbf{\noindent\textit{Discovering previously inaccessible conformational basins.}}  Large-scale resources such as AFDB and ESMAtlas have dramatically expanded structural coverage of protein sequence space \cite{RN63,RN64,RN65}. Yet most predicted structures cluster within previously observed fold space, and only a small fraction (approximately 4\%) appear to represent genuinely novel folds\cite{RN65}. Likewise, large-scale analyses of fold-switching proteins demonstrate that current predictors frequently fail to identify alternative conformations outside their training distributions \cite{RN37}. State-space prediction offers the possibility of discovering previously inaccessible conformational basins, rather than simply assigning sequences to known structural classes.
\newline

\noindent\textbf{\textit{Predicting responses to perturbation.}} A central advantage of state-space prediction is the ability to predict how mutations and environmental changes reshape conformational equilibria. Recent adversarial mutation studies illustrate the limitations of current models: AlphaFold3 predictions often remain nearly unchanged despite extensive sequence perturbations, suggesting a form of nonphysical robustness that fails to capture experimentally observed responses \cite{RN66}. Even deletion of nearly 40\% of the MSA frequently fails to induce expected structural changes. Although ESMFold exhibits greater sensitivity, both models remain more strongly influenced by training-set similarity than by physical plausibility. These observations highlight the need for predictors that model how perturbations redistribute conformational populations rather than simply maintaining coordinate similarity.
\newline

\noindent \textbf{\textit{Conditioning prediction on biological context}} Biological context provides critical information that cannot be inferred from sequence alone. Experimental measurements such as DEER spectroscopy and crosslinking mass spectrometry constrain conformational sampling, enabling methods such as DEERFold and AlphaLink2 to recover experimentally supported conformational ensembles and improve predictions of challenging protein assemblies \cite{RN67,RN68,RN69,RN43,RN54}. Likewise, alternative splicing provides a genetically encoded source of structural variation capable of inserting, deleting, or reorganizing structural elements, generating conformational states absent from current training data \cite{RN70,RN71}. More generally, incorporating environmental conditions, experimental restraints, and intrinsic sequence variation transforms structure prediction into a context-conditioned inference problem in which the relevant prediction depends not only on sequence but also on the biological conditions under which that sequence functions.

\section{Best practices for use of current predictors} 

\subsection{Alternative state sampling}
\begin{figure}
    \centering
    \includegraphics[width=1.0\linewidth]{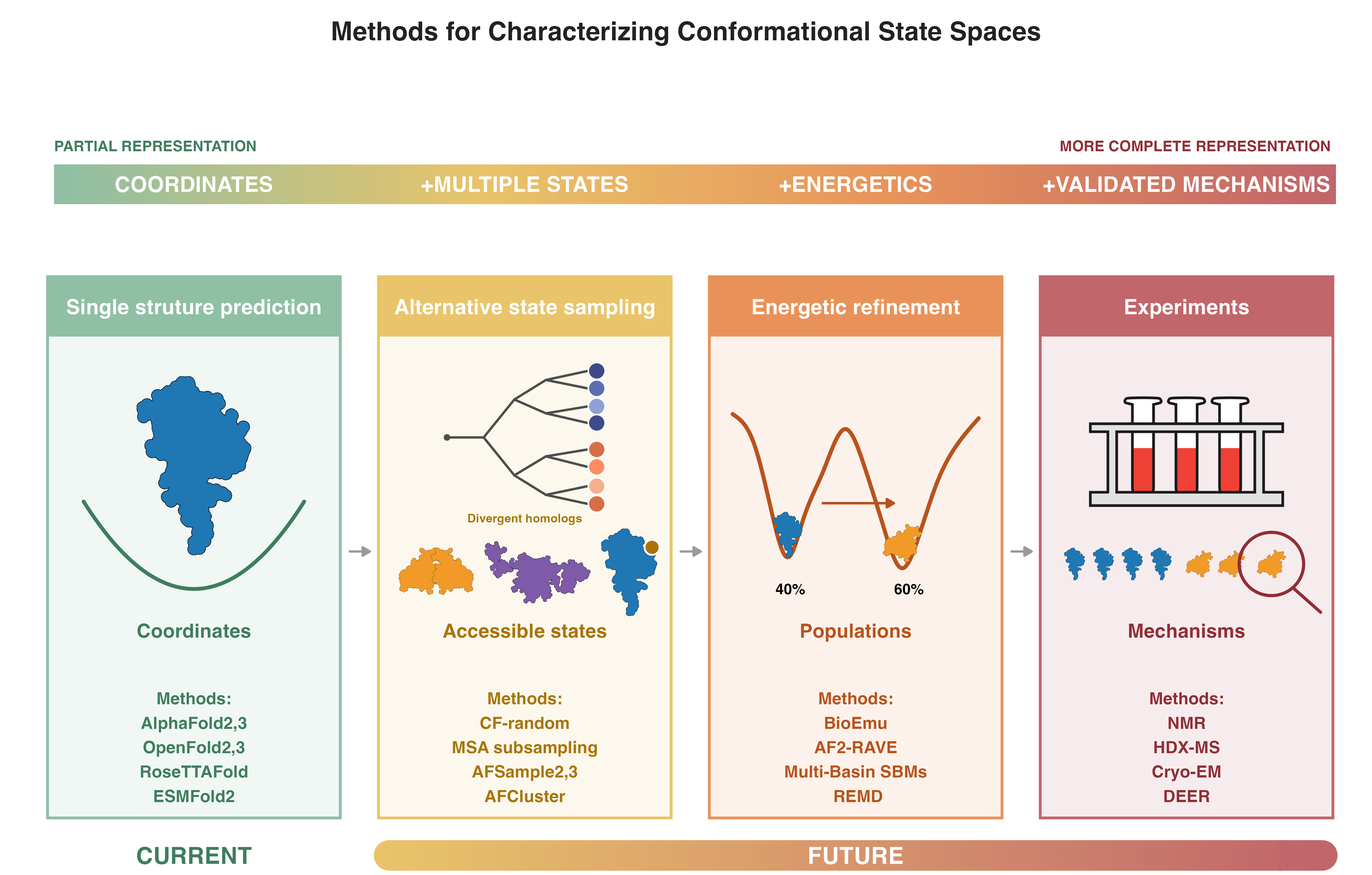}
  \caption{\textbf{Current methods recover complementary components of conformational state space.} Current computational and experimental approaches provide progressively richer representations of conformational state spaces, but no single method yet recovers the complete biological landscape.  \textbf{(1) Single structure prediction} identifies a dominant conformation with high coordinate accuracy but provides little information about alternative states. \textbf{(2) Alternative state sampling} generates candidate accessible conformations using approaches such as MSA perturbation or stochastic sampling, without explicitly estimating their thermodynamic populations.  \textbf{(3) Energetic refinement} combines molecular dynamics, enhanced sampling, and hybrid AI–physics approaches to estimate state populations, free-energy differences, and transition pathways. \textbf{(4) Experiments} provide the ultimate validation of conformational mechanisms by measuring state populations, dynamics, and function.  Together, these complementary approaches recover progressively richer components of conformational state space—coordinates, accessible states, populations, and mechanisms—but integrating these components into a unified state-space predictor remains the central challenge.}
    \label{fig:fig3}
\end{figure}

Current methods for multi-state prediction primarily generate candidate conformational states without explicitly estimating their thermodynamic populations or kinetic accessibility. Rather than predicting a single dominant structure, these approaches can reveal alternative conformations by perturbing AlphaFold-class predictors through strategies such as masking conservation profiles, MSA subsampling, or stochastic sampling \cite{RN45,RN47,RN46}. Early methods adapted AlphaFold2, which generally outperforms AlphaFold3 for alternative-state prediction\cite{RN37}. Because AlphaFold2 behaves almost deterministically for a fixed multiple sequence alignment (MSA), del Alamo \textit{et al.} demonstrated that stochastic MSA subsampling could expose conformations otherwise suppressed by the dominant MSA signal \cite{RN45}. More recently, methods based on local energetic frustration have identified alternative conformations by clustering sequences according to their average frustration, successfully recovering rigid-body domain motions and fold-switching transitions \cite{RN102}. Among AlphaFold2-based approaches, CF-random \cite{RN46} outperformed AFSample2 \cite{kalakoti2025afsample2}, AF-cluster \cite{RN47}, and SPEACH-AF \cite{stein2022speach_af} in predicting domain reorientations, local conformational changes, and fold switching, and was subsequently applied to large-scale prediction of fold-switching proteins in the \textit{E.~coli} proteome, suggesting that up to 5\% of proteins may undergo fold switching. Overall, these methods are most effective when biochemical, biophysical, or evolutionary evidence already suggests the existence of alternative conformations and the goal is to recover a specific non-dominant state \cite{RN48}. While they produce valuable structural hypotheses, they ultimately loosen the single-state objective rather than redefining it.  Collectively, these methods can sometimes recover accessible conformational states, but not their energetic organization or biological relevance (Figure \ref{fig:fig3}).

A second class of methods instead modifies the model architecture or training objective to generate diverse conformations directly from sequence. AlphaFlow  \cite{volk2023alphaflow} exemplifies this approach by modeling conformational fluctuations around an input structure. It successfully recovered experimentally observed minor states of human pro-interleukin-18 (pro-IL-18) when initialized from the experimentally determined structure \cite{RN114}. This example highlights an important limitation shared by many sampling-based methods: the quality of the sampled ensemble depends fundamentally on the correctness of the starting structure. By contrast, AlphaFold2, AlphaFold3, and BioEmu (\cite{RN104}, discussed later) generated starting conformations that differed from the experimentally observed pro-IL-18 structure by more than 20 Å, preventing subsequent sampling from recovering the experimentally observed ensemble \cite{RN61,RN114}. Encouragingly, recent work by Clore and colleagues demonstrated that targeted MSA editing enables both AlphaFold2 and AlphaFold3 to recover a pro-IL-18 conformation similar to experiment without prior structural knowledge \cite{Clore2025.12.30.697132}, suggesting that improved initialization may substantially enhance downstream conformational sampling.

Meanwhile, the confidence metrics of all these methods require careful interpretation. Metrics such as pLDDT and pTM quantify internal geometric consistency rather than thermodynamic stability or kinetic accessibility. Consequently, experimentally validated alternative conformations may receive low confidence, whereas highly confident predictions can simply reflect agreement with dominant evolutionary priors or training data rather than biologically populated states \cite{RN36}. Likewise, mutations that experimentally shift conformational equilibria often produce little visible structural change, indicating that coordinate similarity alone cannot be interpreted as evidence that the underlying conformational landscape is unchanged \cite{Masters2025, kim2025large}. Consequently, these methods are best viewed as generators of candidate conformational states rather than predictors of conformational state spaces. Recovering the energetic organization of those states requires additional physical modeling, discussed next.

\subsection{Energetic refinement}

Sampling alternative conformations represents only the first step toward state-space prediction. Biological function depends not only on which conformations are accessible but also on their relative populations, energetic stability, and kinetic connectivity. Estimating these properties remains substantially more difficult than generating structurally plausible alternative states. While recent deep-learning methods such as BioEmu \cite{RN104} attempt to approximate equilibrium distributions directly, MD simulations remain the principal physics-based approach for estimating free-energy landscapes and transition pathways. Increasingly, hybrid DL-MD methods combine the efficient conformational sampling of deep learning with the physical rigor of MD, providing complementary strategies for reconstructing conformational state spaces (Figure \ref{fig:fig3}).

\subsubsection{Physics-based estimates of conformational landscapes}
Prior to deep learning, computational exploration of conformational landscapes was almost exclusively done using molecular dynamics simulations initiated from experimentally determined structures. As unbiased molecular dynamics simulations are quite expensive to run and slow to sample conformational space; many approaches have been designed to promote effective sampling. Enhanced sampling approaches such as replica-exchange MD run several simulations in parallel at different temperatures and periodically exchange runs to the other temperatures to promote state-space sampling \cite{RN100}. Recognizing that exhaustive sampling is intractable, goal-oriented adaptive sampling strategies such as FAST seek to manage exploration-exploitation trade-offs to maximize the chance that relevant regions of conformational space are explored \cite{zimmerman_fast_2015}. Other simulation strategies, such as targeted MD simulations, apply an external potential to promote structural interconversion between two states. Targeted MD simulations coupled with Markov State Models have been recently used to provide both estimates of the transition kinetics between states as well as the basin connectivity \cite{RN101}.  Finally, structure-based models further reduce computational cost by encoding native contact maps as attractive interactions \cite{RN86}. For proteins undergoing large conformational rearrangements or fold switching, dual-basin structure-based models merge the contact maps of multiple experimentally determined states, allowing efficient reconstruction of transition pathways and free-energy landscapes \cite{RN87,RN95}. These approaches have successfully characterized conformational transitions in adenylate kinase \cite{RN95}, prion protein misfolding \cite{RN96}, and fold-switching proteins including XCL1 \cite{RN97}, KaiB \cite{RN83}, Mad2 \cite{RN98} and RfaH \cite{RN99}. Done correctly, molecular dynamics simulations can be an excellent training source for learning predictors, such as recent efforts to predict the probabilities of forming cryptic pockets \cite{meller_predicting_2023}, as well as the residues which are allosterically linked to cryptic pockets \cite{zhang_ae-pocketminer_2026}.

A major limitation of MD, however, is its dependence on the starting structure \cite{RN127}. Simulations initiated from inaccurate structural models may require microseconds of sampling simply to relax toward experimentally observed conformations. Consequently, the quality of conformational landscape reconstruction depends critically on both the accuracy of the initial model and the environmental conditions under which simulations are performed. 

\subsubsection{Deep learning approximations to energetic weighting}
Recent deep-learning methods have begun addressing energetic weighting directly. Distributional Graphormer (DiG) attempted to learn equilibrium conformational distributions using energy-based training objectives \cite{Zheng2024}, although independent benchmarking has questioned its ability to reproduce realistic ensemble behavior \cite{Aranganathan2026}.  BioEmu represents a more substantial advance, combining AlphaFold Database predictions, molecular dynamics trajectories, and experimental stability measurements to generate approximate Boltzmann-weighted ensembles \cite{RN104}. Unlike purely geometric predictors, it was explicitly trained to reproduce equilibrium observables and was benchmarked not only on near-native structure generation but also on folding free energies and systems with known alternative conformational states, such as cryptic-pocket formation, local unfolding, and domain rearrangements. This makes BioEmu a more sophisticated attempt to move from single-structure prediction toward sequence-conditioned ensemble prediction, though its outputs remain learned approximations to equilibrium distributions rather than direct physical simulations---and, as with pro-IL-18, an incorrect starting structure can still propagate to an inaccurate predicted ensemble \cite{RN114}.

\subsubsection{Hybrid DL-MD methods}
Hybrid methods attempt to combine the complementary strengths of deep learning and molecular dynamics. Deep-learning predictors rapidly identify candidate conformational basins that would otherwise require prohibitively long simulations to discover, while MD evaluates their energetic stability and transition behavior. AlphaFold2-RAVE (AF2-RAVE) exemplifies this strategy by using AlphaFold-generated structures as initial seeds for short unbiased simulations, followed by enhanced sampling to estimate conformational free energies and state populations \cite{RN51, Teng2025}. It utilizes structures generated by DL methods, often reduced-MSA AF2 predictions as structural seeds, and short unbiased MD from those seed structures to explore their vicinity in the conformational landscape. These simulations are gathered to learn a representation of the conformational landscape, where further enhanced sampling simulations can determine relative weights of interesting states and sample more states \cite{RN50}. AF2-RAVE has been used to study the conformational stability for the conserved DFG motif in kinases \cite{RN52}, sidechain-level metastable states of ligand binding pockets for conformation selective docking \cite{RN53}, and accelerated conformational sampling with no prior knowledge \cite{Teng2025}. Similar approaches have been used to accelerate the discovery of cryptic pockets \cite{meller_accelerating_2023}.

BioEmu-generated ensembles have also recently been used as priors for Markov state modeling of kinase conformational landscapes. In fact, this has been recently explored by performing BioEmu-seeded short MD simulations coupled to Markov State models to explore conformational ensembles of serine-threonine kinases, showing good performance in reliably mapping metastable states for these proteins, but failing in other cases such as glycine transporter 1 (GlyT1) and plasmepsin-II (PlmII), proteins in which side chain heterogeneity governs their dynamics \cite{RN105}.  By decoupling conformational discovery from energetic refinement, these hybrid approaches substantially improve sampling efficiency while preserving the physical interpretation of conformational populations. 

\subsubsection{Predicting environmentally conditioned conformational ensembles}

Protein conformations are determined not only by sequence but also by biological context. Ligands, ions, nucleic acids, post-translational modifications, pH, membrane composition, and binding partners can reshape conformational landscapes by stabilizing low-population states or inducing new conformations. Predicting environmentally conditioned conformational ensembles therefore requires models that explicitly incorporate these contextual variables rather than treating proteins as isolated molecules.

Current co-folding methods represent the first generation of context-aware predictors. Rather than predicting an isolated protein and modeling binding afterward, they jointly infer proteins and their interacting molecules. NeuralPLexer \cite{Qiao2024} introduced this end-to-end framework for protein–ligand complexes, followed by RoseTTAFold All-Atom \cite{Krishna2024} and AlphaFold3 \cite{Abramson2024}, which progressively expanded the supported chemical space to include nucleic acids, ions, post-translational modifications, and other biomolecular partners. Because binding frequently shifts conformational equilibria, these models naturally recover context-dependent structural states inaccessible to sequence-only predictors.

Despite these advances, environmentally conditioned prediction remains incomplete. Adversarial studies have shown that co-folding models often predict nearly identical ligand poses even after mutations that eliminate or sterically occlude binding sites, suggesting that current models remain strongly influenced by training-set biases rather than accurately modeling how perturbations reshape conformational landscapes \cite{Masters2025, yu2026bias}. Moreover, many biologically important variables—including pH, ionic composition, temperature, membrane composition, alternative splicing, and macromolecular interactions—remain absent or only indirectly represented. Consequently, current co-folding methods should be viewed as an important first step toward context-aware state-space prediction rather than a complete solution.

Ultimately, environmentally conditioned prediction must be integrated with alternative-state generation and energetic modeling to infer conformational state spaces as functions of both sequence and biological context. Together, these components define the next generation of state-space predictors.

\subsection{Experiments}

Despite these advances, no single computational strategy yet provides a complete description of conformational state spaces. Ensemble-sensitive experimental techniques—including NMR, HDX-MS, SAXS, cryo-EM classification, single-molecule methods, and mutational analyses—remain essential for validating computationally predicted landscapes and characterizing them independently. Structures generated by DL are best seen as hypotheses about potential basins within a conformational landscape. Combining simulation-derived ensembles with experimental observations helps distinguish biologically meaningful states from sampling artifacts \cite{RN55, miller_accounting_2024}. Likewise, expanding FAIR repositories of molecular dynamics trajectories will provide increasingly valuable training data for future landscape-aware predictors \cite{RN129}. For example, BioEmu was trained on almost 100 milliseconds of publicly available all-atom molecular dynamics simulations \cite{RN104}. For MD simulations, careful consideration must be given to force-field dependence, sampling biases, and the enormous computational cost of generating sufficiently converged simulations. Even the longest unbiased MD simulation trajectory from DESRES barely captures one or two relevant slow conformational events for proteins such as kinases \cite{shan2011does}. Thus, one runs the risk of either having MD trajectories that are too short and unintentionally biased due to simulation setup, or having an explosion of perhaps petabytes or more of MD trajectories. Careful use of enhanced sampling in generating such a database of MD trajectories might be a compromise, especially if different sampling schemes yield similar ensembles. Ultimately, deep learning, molecular dynamics, and experiment should be viewed as complementary components of a single iterative framework: deep learning proposes candidate states, physics estimates their energetic organization, and experiments establish their biological relevance. Until predictors directly recover complete conformational landscapes, ensemble-aware interpretation of computational predictions will remain essential (Figure \ref{fig:fig3}).

\section{Toward State-Space Prediction}
\begin{figure}
    \centering
    \includegraphics[width=1.0\linewidth]{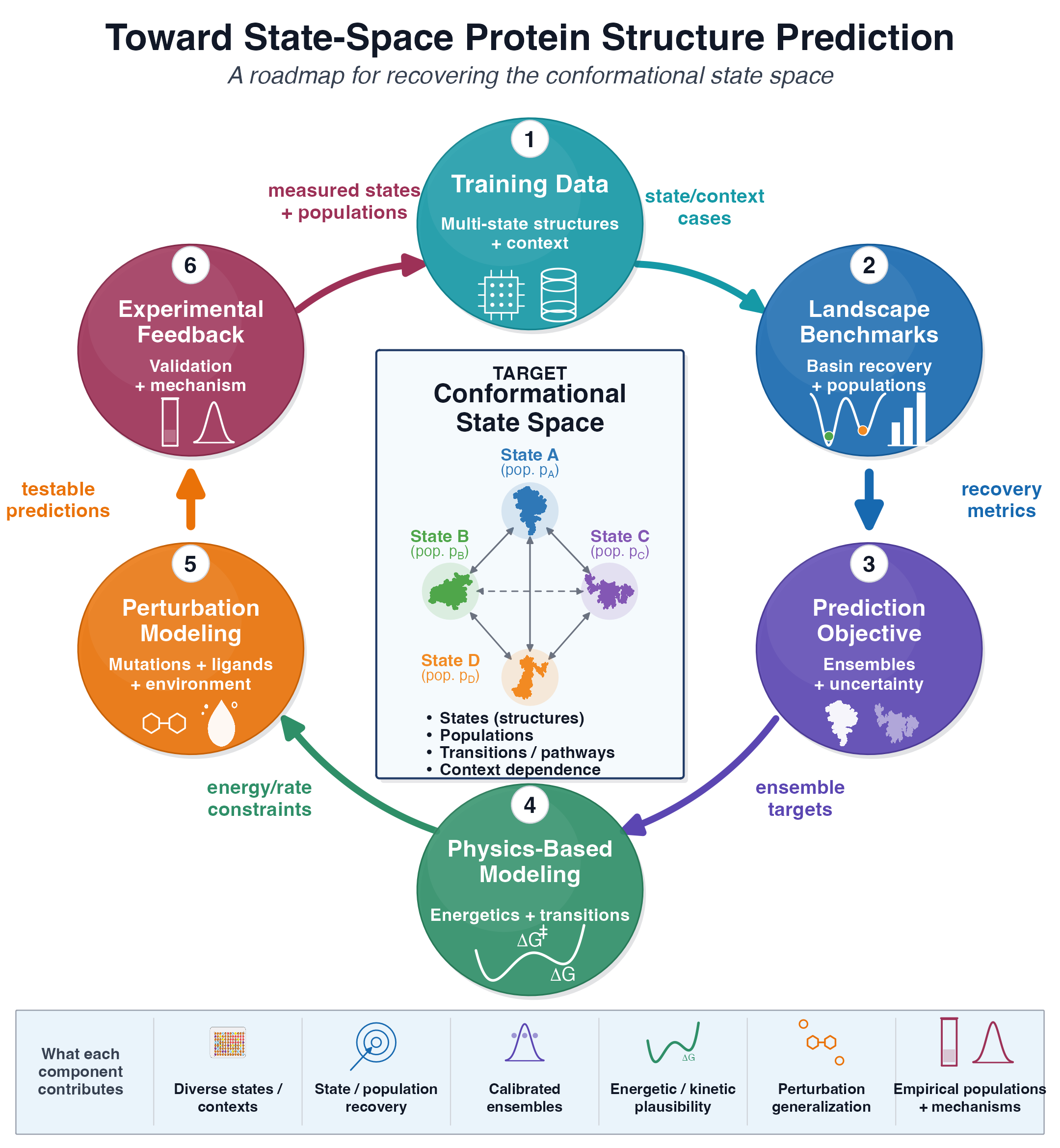}
     \caption{\textbf{A roadmap for state-space prediction.} The field's target is the conformational state space of a protein (center): the set of accessible states, their relative populations, the transitions or pathways connecting them, and their dependence on biological context. Reaching this target requires progress across six interlocking components, arranged as a cycle. \textbf{(1) Training data} — multi-state structures collected under diverse contextual conditions — \textbf{teaches (2) landscape benchmarks}, which provide quantitative tests of basin and state recovery. These benchmarks \textbf{evaluate (3) the prediction objective} itself, reframing model outputs as ensembles with calibrated uncertainty rather than single coordinates. That objective in turn \textbf{defines (4) physics-based modeling} of the energetics and transition rates ($\Delta G$, $\Delta G^\ddagger$) linking states, which \textbf{constrains (5) perturbation modeling} — predicting how mutations, ligands, pH, or temperature reshape the landscape and generalize to new conditions. Perturbation models \textbf{generate predictions} that motivate \textbf{(6) experimental feedback}\textbf{,} supplying empirical populations and mechanistic ground truth that \textbf{validate and inform} the training data, closing the loop. Rather than a linear pipeline, this cycle emphasizes that improvements in any one component both depend on and drive improvements in the others.}
    \label{fig:fig4}
\end{figure}

Current methods increasingly recover individual components of conformational state spaces, including alternative conformations, approximate energetic weighting, and environmentally conditioned structures. The next challenge is to integrate these capabilities into predictors that treat conformational state spaces as the primary prediction target. One major opportunity is improvement of confidence metrics. Current metrics primarily assess structural self-consistency based on learned priors and training data, rather than thermodynamic stability or basin occupancy. Consequently, structures with high confidence may reflect dominant training-set conformations without indicating their energetic plausibility \cite{RN48,RN49}. Landscape-aware prediction, therefore, requires methods that estimate basin stability and distinguish metastable conformations from transient or artifactual states. Achieving this likely requires closer integration between generative DL approaches and physics-based methods, such as enhanced sampling, free energy estimation, and hybrid DL-molecular dynamics frameworks capable of evaluating the energetic properties of candidate states (Figure \ref{fig:fig4}).

\subsection{Conditioning on Biological Context}

Basin stability is only one property of the landscape that predictors must recover. Many current structure prediction methods are built to predict one or a small ranked set of coordinate models $\{X\}$ for a sequence $S$. This objective reflects the training data: stable, experimentally accessible conformations deposited in the PDB, with only a few entries explicitly showing alternative states. Even when multiple conformations are present for the same protein in the training set, they are usually treated as separate training examples rather than as parts of a shared conformational landscape. Basin identity, transition architecture, and mutational rewiring are further properties of the landscape itself. But the landscape is not fixed: it is reshaped by the conditions $C$ under which a protein exists, and next-generation predictors must treat these conditions as first-class inputs rather than post hoc corrections. Three sources of context are especially informative, because each constrains $p(X\mid S, C)$ in a distinct way: direct experimental measurement, environmental state, and intrinsic sequence variation arising from splicing.

\paragraph{Experimental constraints.} Sparse experimental measurements can convert an underdetermined sampling problem into a constrained inference problem. DEERFold incorporates distance distributions from Double Electron--Electron Resonance (DEER) spectroscopy directly into the network, guiding sampling toward conformations consistent with experimentally observed ensembles \cite{RN67}, and has been used to recover conformational switching in membrane transporters and soluble proteins that a sequence-only model would not surface. AlphaLink2 similarly uses crosslinking mass spectrometry restraints to improve predictions of protein complexes, by identifying interfaces, focusing sampling, and improving model selection---particularly for heteromeric assemblies where evolutionary signal from MSAs is weak or absent \cite{RN49,RN58,RN61,RN62}. In both cases, a small number of distance or contact restraints meaningfully narrows the space of plausible basins, illustrating that context-conditioning need not require dense experimental data to be useful.

\paragraph{Alternative splicing as intrinsic context.} Splicing differs from an experimental restraint in kind, not just degree: it is a genetically encoded source of structural variation rather than an external measurement. Rather than introducing small sequence perturbations, splicing can insert, delete, or reorder entire structural elements---including regions within otherwise conserved domains---producing substantial rearrangements and, in some cases, transitions between distinct folds \cite{RN70}. Large-scale analyses show that splice isoforms can alter surface charge, compactness, and the accessibility of regulatory sites such as post-translational modification sites, thereby reshaping interaction networks and downstream function \cite{RN71}. Because isoform expression is often cell-type- and condition-specific, the biologically relevant conformational ensemble for a given transcript depends on more than sequence alone---it depends on the cellular context in which that sequence is expressed. From a modeling standpoint, this makes splicing a particularly demanding test of context-awareness: it can generate structural states entirely absent from training data, which models that assume one canonical sequence per gene are structurally incapable of recovering.

\paragraph{Why conditioning matters for landscape prediction specifically.} Each context source acts on $p(X\mid S, C)$ in a way a single-structure objective cannot represent even in principle: experimental restraints reshape sampling toward in vivo-relevant states rather than the most training-consistent one \cite{RN67}; environmental and binding conditions shift basin occupancy directly; and splicing changes the accessible basin set by altering the effective sequence input itself. A predictor recovering only $X$ has no mechanism for any of these effects, whereas one built around $p(X\mid S, C)$ can absorb all three as conditioning inputs rather than special cases. Complex and interface modeling stands to benefit immediately, since evolutionary signal is weakest exactly where context is most informative—making context-conditioning and multi-state prediction complementary, not competing, extensions of the single-structure paradigm.

\subsection{Evaluation Strategies for Landscape-Aware Models}

Finally, evaluation strategies must align with this expanded objective. Traditional metrics such as RMSD and TM-score reward recovery of a single dominant fold and are poorly suited to assessing multi-state behavior. Landscape-aware evaluation should instead measure recovery of experimentally validated alternative conformations, discrimination between competing basins, and sensitivity to perturbations such as mutation, ligand binding, or the contextual factors described above (experimental restraints, environmental condition, splice isoform). Benchmark datasets must therefore include proteins with known conformational heterogeneity and separate alternative states across training and evaluation splits to ensure genuine generalization \cite{brown2025generalizable}, rather than memorization of a training-set-dominant basin. Recent benchmarking efforts comparing simulation- and AI-based estimates of cryptic pocket probabilities to experimental measurements of these thermodynamic quantities are a good step in this direction \cite{zhang_how_2026,meller_predicting_2023}

Taken together, these requirements define the transition from structure prediction to landscape prediction. Single-structure prediction delivered unprecedented accuracy in coordinate modeling; state-space prediction represents the next step toward understanding the dynamic mechanisms through which proteins encode function. Overcoming this formidable challenge will likely require integrating generative DL with physics-based approaches such as enhanced sampling, free-energy estimation, and hybrid DL-molecular dynamics frameworks that evaluate energetic stability. We highlight how Levy and co-workers have shown that sequence-based Potts Hamiltonian models, trained on kinase-family MSAs, can estimate mutational effects on kinase stability and agree with both experimental thermostability data and MD/FEP-derived $\Delta\Delta G$ values \cite{thakur2024potts}. Extending this idea to kinase inhibitor selectivity, they used Potts statistical energies to quantify the reorganization cost between active and inactive kinase conformations, showing that conformational reorganization contributes substantially to type-II inhibitor binding across the kinome \cite{vardanyan2026results}.  

\section*{Conclusion and Perspective}
Protein structure prediction has achieved a major milestone: accurately identifying a single dominant conformation is no longer the main obstacle. The real challenge now is conceptual. Proteins are dynamic entities whose functions depend on the organization, accessibility, and interconversion of multiple conformational states. As experimental techniques increasingly reveal conformational diversity and context-dependent behavior, the biological object of interest is a landscape, not a static structure. Existing predictive models, designed to find a single state, only reflect a part of this landscape and thus overlook how proteins regulate activity, recognize ligands, and respond to changes. 

The next step requires redefining the prediction goal. Instead of asking \textit{"What is the structure?''}, we should ask \textit{``How is the conformational equilibrium organized, how does it change in response to mutations and changes in the environment?''}. Achieving this involves developing models that combine DL with physical principles, MD simulations and experimental data to predict not only structures but also energetic relationships and transition pathways. This shift offers significant benefits: context-aware predictions will enable us to directly infer mechanisms such as ligand stabilization of specific states, the effects of mutations on conformational populations, and the influence of regulatory signals on protein behavior. This has direct implications for drug discovery, protein design, and systems biology, where function relies on dynamic state changes rather than static structures. While single-structure prediction has largely solved the problem of coordinate accuracy, the next breakthrough will be measured not in angstroms, but in our ability to infer conformational state spaces and the biological mechanisms they encode.

\section*{Code and Data}
All code and data used to generate figures can be found at: \url{https://github.com/DevlinaC/expanding-protein-structure-prediction-figures}.

\section*{Acknowledgements}
This research was supported in part by the Division of Intramural Research at the National Library of Medicine, National Institutes of Health (NIH, LM202011 to L.L.P., GM152085 to G.R.B), the National Agency for Research and Development (ANID) through Fondo Nacional de Desarrollo Científico y Tecnológico (FONDECYT Regular 1240205 to C.A.R-S.; ANID PFCHA 21230688 to C.N-M.), the ANID Millennium Science Initiative Program (ICN17\_022 to C.A.R-S.), the Natural Sciences and Engineering Research Council of Canada (RGPIN-2025-07032 to S.R. and CGS-D to Y.O.R.), the Digital Research Alliance of Canada (S.R.). The contributions of the NIH authors are considered Works of the United States Government. The findings and conclusions presented in this paper are those of the authors and do not necessarily reflect the views of the NIH or the U.S. Department of Health and Human Services. 

\bibliography{references}

@article{RN1,
   author = {Anfinsen, C. B. and Haber, E. and Sela, M. and White, F. H.},
   title = {THE KINETICS OF FORMATION OF NATIVE RIBONUCLEASE DURING OXIDATION OF THE REDUCED POLYPEPTIDE CHAIN},
   journal = {Proceedings of the National Academy of Sciences},
   volume = {47},
   number = {9},
   pages = {1309-1314},
   DOI = {doi:10.1073/pnas.47.9.1309},
   url = {https://www.pnas.org/doi/abs/10.1073/pnas.47.9.1309},
   year = {1961},
   type = {Journal Article}
}

@article{RN11,
   author = {Wolynes, Peter G. and Onuchic, Jose N. and Thirumalai, D.},
   title = {Navigating the Folding Routes},
   journal = {Science},
   volume = {267},
   number = {5204},
   pages = {1619-1620},
   DOI = {10.1126/science.7886447},
   url = {https://doi.org/10.1126/science.7886447},
   year = {1995},
   type = {Journal Article}
}

@article{RN9,
   author = {Dill, Ken A. and Chan, Hue Sun},
   title = {From {Levinthal} to pathways to funnels},
   journal = {Nature Structural Biology},
   volume = {4},
   number = {1},
   pages = {10-19},
   ISSN = {1545-9985},
   DOI = {10.1038/nsb0197-10},
   url = {https://doi.org/10.1038/nsb0197-10},
   year = {1997},
   type = {Journal Article}
}

@article{RN10,
   author = {Dobson, Christopher M. and Karplus, Martin},
   title = {The fundamentals of protein folding: bringing together theory and experiment},
   journal = {Current Opinion in Structural Biology},
   volume = {9},
   number = {1},
   pages = {92-101},
   ISSN = {0959-440X},
   DOI = {https://doi.org/10.1016/S0959-440X(99)80012-8},
   url = {https://www.sciencedirect.com/science/article/pii/S0959440X99800128},
   year = {1999},
   type = {Journal Article}
}

@article{RN56,
   author = {Rost, Burkhard},
   title = {Twilight zone of protein sequence alignments},
   journal = {Protein Engineering},
   volume = {12},
   number = {2},
   pages = {85-94},
   ISSN = {0269-2139},
   DOI = {10.1093/protein/12.2.85},
   url = {https://doi.org/10.1093/protein/12.2.85},
   year = {1999},
   type = {Journal Article}
}

@article{RN12,
   author = {Yernool, Dinesh and Boudker, Olga and Jin, Yan and Gouaux, Eric},
   title = {Structure of a glutamate transporter homologue from {Pyrococcus} horikoshii},
   journal = {Nature},
   volume = {431},
   number = {7010},
   pages = {811-818},
   ISSN = {1476-4687},
   DOI = {10.1038/nature03018},
   url = {https://doi.org/10.1038/nature03018},
   year = {2004},
   type = {Journal Article}
}

@article{RN27,
   author = {Mittermaier, A. and Kay, L. E.},
   title = {New tools provide new insights in {NMR} studies of protein dynamics},
   journal = {Science},
   volume = {312},
   number = {5771},
   pages = {224-8},
   ISSN = {0036-8075},
   DOI = {10.1126/science.1124964},
   year = {2006},
   type = {Journal Article}
}

@article{RN13,
   author = {Boudker, Olga and Ryan, Renae M. and Yernool, Dinesh and Shimamoto, Keiko and Gouaux, Eric},
   title = {Coupling substrate and ion binding to extracellular gate of a sodium-dependent aspartate transporter},
   journal = {Nature},
   volume = {445},
   number = {7126},
   pages = {387-393},
   ISSN = {1476-4687},
   DOI = {10.1038/nature05455},
   url = {https://doi.org/10.1038/nature05455},
   year = {2007},
   type = {Journal Article}
}

@article{RN70,
   author = {Birzele, F. and Csaba, G. and Zimmer, R.},
   title = {Alternative splicing and protein structure evolution},
   journal = {Nucleic Acids Res},
   volume = {36},
   number = {2},
   pages = {550-8},
   ISSN = {0305-1048 (Print)
0305-1048},
   DOI = {10.1093/nar/gkm1054},
   year = {2008},
   type = {Journal Article}
}

@article{RN4,
   author = {Boehr, David D. and Nussinov, Ruth and Wright, Peter E.},
   title = {The role of dynamic conformational ensembles in biomolecular recognition},
   journal = {Nature Chemical Biology},
   volume = {5},
   number = {11},
   pages = {789-796},
   ISSN = {1552-4469},
   DOI = {10.1038/nchembio.232},
   url = {https://doi.org/10.1038/nchembio.232},
   year = {2009},
   type = {Journal Article}
}

@article{RN30,
   author = {Konermann, Lars and Pan, Jingxi and Liu, Yu-Hong},
   title = {Hydrogen exchange mass spectrometry for studying protein structure and dynamics},
   journal = {Chemical Society Reviews},
   volume = {40},
   number = {3},
   pages = {1224-1234},
   ISSN = {0306-0012},
   DOI = {10.1039/C0CS00113A},
   url = {http://dx.doi.org/10.1039/C0CS00113A},
   year = {2011},
   type = {Journal Article}
}

@article{RN28,
   author = {Rosato, A. and Aramini, J. M. and Arrowsmith, C. and Bagaria, A. and Baker, D. and Cavalli, A. and Doreleijers, J. F. and Eletsky, A. and Giachetti, A. and Guerry, P. and Gutmanas, A. and Güntert, P. and He, Y. and Herrmann, T. and Huang, Y. J. and Jaravine, V. and Jonker, H. R. and Kennedy, M. A. and Lange, O. F. and Liu, G. and Malliavin, T. E. and Mani, R. and Mao, B. and Montelione, G. T. and Nilges, M. and Rossi, P. and van der Schot, G. and Schwalbe, H. and Szyperski, T. A. and Vendruscolo, M. and Vernon, R. and Vranken, W. F. and Vries, Sd and Vuister, G. W. and Wu, B. and Yang, Y. and Bonvin, A. M.},
   title = {Blind testing of routine, fully automated determination of protein structures from {NMR} data},
   journal = {Structure},
   volume = {20},
   number = {2},
   pages = {227-36},
   ISSN = {0969-2126 (Print)
0969-2126},
   DOI = {10.1016/j.str.2012.01.002},
   year = {2012},
   type = {Journal Article}
}

@article{RN32,
   author = {Stigler, Johannes and Rief, Matthias},
   title = {Hidden Markov Analysis of Trajectories in Single-Molecule Experiments and the Effects of Missed Events},
   journal = {ChemPhysChem},
   volume = {13},
   number = {4},
   pages = {1079-1086},
   ISSN = {1439-4235},
   DOI = {https://doi.org/10.1002/cphc.201100814},
   url = {https://doi.org/10.1002/cphc.201100814},
   year = {2012},
   type = {Journal Article}
}

@article{RN31,
   author = {Lerner, E. and Cordes, T. and Ingargiola, A. and Alhadid, Y. and Chung, S. and Michalet, X. and Weiss, S.},
   title = {Toward dynamic structural biology: Two decades of single-molecule {Förster} resonance energy transfer},
   journal = {Science},
   volume = {359},
   number = {6373},
   ISSN = {0036-8075 (Print)
0036-8075},
   DOI = {10.1126/science.aan1133},
   year = {2018},
   type = {Journal Article}
}

@article{RN16,
   author = {Porter, Justin R. and Moeder, Katelyn E. and Sibbald, Carrie A. and Zimmerman, Maxwell I. and Hart, Kathryn M. and Greenberg, Michael J. and Bowman, Gregory R.},
   title = {Cooperative Changes in Solvent Exposure Identify Cryptic Pockets, Switches, and Allosteric Coupling},
   journal = {Biophysical Journal},
   volume = {116},
   number = {5},
   pages = {818-830},
   ISSN = {0006-3495},
   DOI = {https://doi.org/10.1016/j.bpj.2018.11.3144},
   url = {https://www.sciencedirect.com/science/article/pii/S0006349519300530},
   year = {2019},
   type = {Journal Article}
}

@article{RN20,
   author = {Guttman, Miklos and Padte, Neal N. and Huang, Yaoxing and Yu, Jian and Rocklin, Gabriel J. and Weitzner, Brian D. and Scian, Michele and Ho, David D. and Lee, Kelly K.},
   title = {The influence of proline isomerization on potency and stability of anti-{HIV} antibody {10E8}},
   journal = {Scientific Reports},
   volume = {10},
   number = {1},
   pages = {14313},
   ISSN = {2045-2322},
   DOI = {10.1038/s41598-020-71184-7},
   url = {https://doi.org/10.1038/s41598-020-71184-7},
   year = {2020},
   type = {Journal Article}
}

@article{RN38,
   author = {Jumper, John and Evans, Richard and Pritzel, Alexander and Green, Tim and Figurnov, Michael and Ronneberger, Olaf and Tunyasuvunakool, Kathryn and Bates, Russ and Žídek, Augustin and Potapenko, Anna and Bridgland, Alex and Meyer, Clemens and Kohl, Simon A. A. and Ballard, Andrew J. and Cowie, Andrew and Romera-Paredes, Bernardino and Nikolov, Stanislav and Jain, Rishub and Adler, Jonas and Back, Trevor and Petersen, Stig and Reiman, David and Clancy, Ellen and Zielinski, Michal and Steinegger, Martin and Pacholska, Michalina and Berghammer, Tamas and Bodenstein, Sebastian and Silver, David and Vinyals, Oriol and Senior, Andrew W. and Kavukcuoglu, Koray and Kohli, Pushmeet and Hassabis, Demis},
   title = {Highly accurate protein structure prediction with {AlphaFold}},
   journal = {Nature},
   volume = {596},
   number = {7873},
   pages = {583-589},
   ISSN = {1476-4687},
   DOI = {10.1038/s41586-021-03819-2},
   url = {https://doi.org/10.1038/s41586-021-03819-2},
   year = {2021},
   type = {Journal Article}
}

@article{RN50,
   author = {Wang, Dedi and Tiwary, Pratyush},
   title = {State predictive information bottleneck},
   journal = {The Journal of Chemical Physics},
   volume = {154},
   number = {13},
   ISSN = {0021-9606},
   DOI = {10.1063/5.0038198},
   url = {https://doi.org/10.1063/5.0038198},
   year = {2021},
   type = {Journal Article}
}

@article{RN15,
   author = {Cruz, Matthew A. and Frederick, Thomas E. and Mallimadugula, Upasana L. and Singh, Sukrit and Vithani, Neha and Zimmerman, Maxwell I. and Porter, Justin R. and Moeder, Katelyn E. and Amarasinghe, Gaya K. and Bowman, Gregory R.},
   title = {A cryptic pocket in {Ebola} {VP35} allosterically controls {RNA} binding},
   journal = {Nature Communications},
   volume = {13},
   number = {1},
   pages = {2269},
   ISSN = {2041-1723},
   DOI = {10.1038/s41467-022-29927-9},
   url = {https://doi.org/10.1038/s41467-022-29927-9},
   year = {2022},
   type = {Journal Article}
}

@article{RN45,
   author = {del Alamo, Diego and Sala, Davide and McHaourab, Hassane S. and Meiler, Jens},
   title = {Sampling alternative conformational states of transporters and receptors with {AlphaFold2}},
   journal = {eLife},
   volume = {11},
   pages = {e75751},
   ISSN = {2050-084X},
   DOI = {10.7554/eLife.75751},
   url = {https://doi.org/10.7554/eLife.75751},
   year = {2022},
   type = {Journal Article}
}

@article{RN69,
   author = {Li, Ziyao and Liu, Xuyang and Chen, Weijie and Shen, Fan and Bi, Hangrui and Ke, Guolin and Zhang, Linfeng},
   title = {{Uni-Fold}: An Open-Source Platform for Developing Protein Folding Models beyond {AlphaFold}},
   journal = {bioRxiv},
   pages = {2022.08.04.502811},
   DOI = {10.1101/2022.08.04.502811},
   url = {https://www.biorxiv.org/content/biorxiv/early/2022/08/30/2022.08.04.502811.full.pdf},
   year = {2022},
   type = {Journal Article}
}

@article{RN63,
   author = {Varadi, M. and Anyango, S. and Deshpande, M. and Nair, S. and Natassia, C. and Yordanova, G. and Yuan, D. and Stroe, O. and Wood, G. and Laydon, A. and Žídek, A. and Green, T. and Tunyasuvunakool, K. and Petersen, S. and Jumper, J. and Clancy, E. and Green, R. and Vora, A. and Lutfi, M. and Figurnov, M. and Cowie, A. and Hobbs, N. and Kohli, P. and Kleywegt, G. and Birney, E. and Hassabis, D. and Velankar, S.},
   title = {{AlphaFold} {Protein} {Structure} {Database}: massively expanding the structural coverage of protein-sequence space with high-accuracy models},
   journal = {Nucleic Acids Res},
   volume = {50},
   number = {D1},
   pages = {D439-d444},
   ISSN = {0305-1048 (Print)
0305-1048},
   DOI = {10.1093/nar/gkab1061},
   year = {2022},
   type = {Journal Article}
}

@article{RN65,
   author = {Barrio-Hernandez, Inigo and Yeo, Jingi and Jänes, Jürgen and Mirdita, Milot and Gilchrist, Cameron L. M. and Wein, Tanita and Varadi, Mihaly and Velankar, Sameer and Beltrao, Pedro and Steinegger, Martin},
   title = {Clustering predicted structures at the scale of the known protein universe},
   journal = {Nature},
   volume = {622},
   number = {7983},
   pages = {637-645},
   ISSN = {1476-4687},
   DOI = {10.1038/s41586-023-06510-w},
   url = {https://doi.org/10.1038/s41586-023-06510-w},
   year = {2023},
   type = {Journal Article}
}

@article{RN64,
   author = {Burke, D. F. and Bryant, P. and Barrio-Hernandez, I. and Memon, D. and Pozzati, G. and Shenoy, A. and Zhu, W. and Dunham, A. S. and Albanese, P. and Keller, A. and Scheltema, R. A. and Bruce, J. E. and Leitner, A. and Kundrotas, P. and Beltrao, P. and Elofsson, A.},
   title = {Towards a structurally resolved human protein interaction network},
   journal = {Nat Struct Mol Biol},
   volume = {30},
   number = {2},
   pages = {216-225},
   ISSN = {1545-9993 (Print)
1545-9985},
   DOI = {10.1038/s41594-022-00910-8},
   year = {2023},
   type = {Journal Article}
}

@article{RN7,
   author = {Chakravarty, Devlina and Schafer, Joseph W. and Porter, Lauren L.},
   title = {Distinguishing features of fold-switching proteins},
   journal = {Protein Science},
   volume = {32},
   number = {3},
   pages = {e4596},
   ISSN = {0961-8368},
   DOI = {https://doi.org/10.1002/pro.4596},
   url = {https://doi.org/10.1002/pro.4596},
   year = {2023},
   type = {Journal Article}
}

@article{RN57,
   author = {Chakravarty, Devlina and Sreenivasan, Shwetha and Swint-Kruse, Liskin and Porter, Lauren L.},
   title = {Identification of a covert evolutionary pathway between two protein folds},
   journal = {Nature Communications},
   volume = {14},
   number = {1},
   pages = {3177},
   ISSN = {2041-1723},
   DOI = {10.1038/s41467-023-38519-0},
   url = {https://doi.org/10.1038/s41467-023-38519-0},
   year = {2023},
   type = {Journal Article}
}

@article{RN60,
   author = {Liu, Shun and Chen, Hua and Yin, Yan and Lu, Defen and Gao, Guoming and Li, Jie and Bai, Xiao-Chen and Zhang, Xuewu},
   title = {Inhibition of {FAM46/TENT5} activity by {BCCIP}$\alpha$ adopting a unique fold},
   journal = {Science Advances},
   volume = {9},
   number = {14},
   pages = {eadf5583},
   DOI = {doi:10.1126/sciadv.adf5583},
   url = {https://www.science.org/doi/abs/10.1126/sciadv.adf5583},
   year = {2023},
   type = {Journal Article}
}

@article{RN59,
   author = {Porter, Lauren L.},
   title = {Fluid protein fold space and its implications},
   journal = {BioEssays},
   volume = {45},
   number = {9},
   pages = {2300057},
   ISSN = {0265-9247},
   DOI = {https://doi.org/10.1002/bies.202300057},
   url = {https://doi.org/10.1002/bies.202300057},
   year = {2023},
   type = {Journal Article}
}

@article{RN21,
   author = {Ruan, Biao and He, Yanan and Chen, Yingwei and Choi, Eun Jung and Chen, Yihong and Motabar, Dana and Solomon, Tsega and Simmerman, Richard and Kauffman, Thomas and Gallagher, D Travis},
   title = {Design and characterization of a protein fold switching network},
   journal = {Nature Communications},
   volume = {14},
   number = {1},
   pages = {431},
   ISSN = {2041-1723},
   DOI = {https://doi.org/10.1038/s41467-023-36065-3},
   url = {https://doi.org/10.1038/s41467-023-36065-3},
   year = {2023},
   type = {Journal Article}
}

@article{RN51,
   author = {Vani, Bodhi P. and Aranganathan, Akashnathan and Wang, Dedi and Tiwary, Pratyush},
   title = {{AlphaFold2-RAVE}: From Sequence to {Boltzmann} Ranking},
   journal = {Journal of Chemical Theory and Computation},
   volume = {19},
   number = {14},
   pages = {4351-4354},
   ISSN = {1549-9618},
   DOI = {10.1021/acs.jctc.3c00290},
   url = {https://doi.org/10.1021/acs.jctc.3c00290},
   year = {2023},
   type = {Journal Article}
}

@article{RN48,
   author = {Agarwal, Vinayak and McShan, Andrew C.},
   title = {The power and pitfalls of {AlphaFold2} for structure prediction beyond rigid globular proteins},
   journal = {Nature Chemical Biology},
   volume = {20},
   number = {8},
   pages = {950-959},
   ISSN = {1552-4469},
   DOI = {10.1038/s41589-024-01638-w},
   url = {https://doi.org/10.1038/s41589-024-01638-w},
   year = {2024},
   type = {Journal Article}
}

@article{RN61,
   author = {Bonin, Jeffrey P. and Aramini, James M. and Dong, Ying and Wu, Hao and Kay, Lewis E.},
   title = {{AlphaFold2} as a replacement for solution {NMR} structure determination of small proteins: Not so fast!},
   journal = {Journal of Magnetic Resonance},
   volume = {364},
   pages = {107725},
   ISSN = {1090-7807},
   DOI = {https://doi.org/10.1016/j.jmr.2024.107725},
   url = {https://www.sciencedirect.com/science/article/pii/S1090780724001095},
   year = {2024},
   type = {Journal Article}
}

@article{RN26,
   author = {Bryant, Patrick and Noé, Frank},
   title = {Structure prediction of alternative protein conformations},
   journal = {Nature Communications},
   volume = {15},
   number = {1},
   pages = {7328},
   ISSN = {2041-1723},
   DOI = {10.1038/s41467-024-51507-2},
   url = {https://doi.org/10.1038/s41467-024-51507-2},
   year = {2024},
   type = {Journal Article}
}

@article{RN37,
   author = {Chakravarty, D. and Schafer, J. W. and Chen, E. A. and Thole, J. F. and Ronish, L. A. and Lee, M. and Porter, L. L.},
   title = {{AlphaFold} predictions of fold-switched conformations are driven by structure memorization},
   journal = {Nat Commun},
   volume = {15},
   number = {1},
   pages = {7296},
   ISSN = {2041-1723},
   DOI = {10.1038/s41467-024-51801-z},
   year = {2024},
   type = {Journal Article}
}

@misc{RN53,
   author = {Gu, Xinyu and Aranganathan, Akashnathan and Tiwary, Pratyush},
   title = {Empowering {AlphaFold2} for protein conformation selective drug discovery with {AlphaFold2-RAVE}},
   publisher = {eLife Sciences Publications, Ltd},
   month = {2024/08/14},
   DOI = {10.7554/elife.99702.2},
   url = {http://dx.doi.org/10.7554/eLife.99702.2},
   year = {2024},
   type = {Generic}
}

@article{RN68,
   author = {Stahl, Kolja and Warneke, Robert and Demann, Lorenz and Bremenkamp, Rica and Hormes, Björn and Brock, Oliver and Stülke, Jörg and Rappsilber, Juri},
   title = {Modelling protein complexes with crosslinking mass spectrometry and deep learning},
   journal = {Nature Communications},
   volume = {15},
   number = {1},
   pages = {7866},
   ISSN = {2041-1723},
   DOI = {10.1038/s41467-024-51771-2},
   url = {https://doi.org/10.1038/s41467-024-51771-2},
   year = {2024},
   type = {Journal Article}
}

@article{RN52,
   author = {Vani, Bodhi P. and Aranganathan, Akashnathan and Tiwary, Pratyush},
   title = {Exploring Kinase {Asp-Phe-Gly} {(DFG)} Loop Conformational Stability with {AlphaFold2-RAVE}},
   journal = {Journal of Chemical Information and Modeling},
   volume = {64},
   number = {7},
   pages = {2789-2797},
   ISSN = {1549-9596},
   DOI = {10.1021/acs.jcim.3c01436},
   url = {https://doi.org/10.1021/acs.jcim.3c01436},
   year = {2024},
   type = {Journal Article}
}

@article{RN47,
   author = {Wayment-Steele, Hannah K. and Ojoawo, Adedolapo and Otten, Renee and Apitz, Julia M. and Pitsawong, Warintra and Hömberger, Marc and Ovchinnikov, Sergey and Colwell, Lucy and Kern, Dorothee},
   title = {Predicting multiple conformations via sequence clustering and {AlphaFold2}},
   journal = {Nature},
   volume = {625},
   number = {7996},
   pages = {832-839},
   ISSN = {1476-4687},
   DOI = {10.1038/s41586-023-06832-9},
   url = {https://doi.org/10.1038/s41586-023-06832-9},
   year = {2024},
   type = {Journal Article}
}

@article{RN22,
   author = {Wayment-Steele, Hannah K and Otten, Renee and Pitsawong, Warintra and Ojoawo, Adedolapo M and Glaser, Andrew and Calderone, Logan A and Kern, Dorothee},
   title = {The conformational landscape of fold-switcher {KaiB} is tuned to the circadian rhythm timescale},
   journal = {Proceedings of the National Academy of Sciences},
   volume = {121},
   number = {45},
   pages = {e2412293121},
   ISSN = {0027-8424},
   doi = {10.1073/pnas.2412293121},
   URL = {https://www.pnas.org/doi/abs/10.1073/pnas.2412293121},
   year = {2024},
   type = {Journal Article}
}

@article{RN58,
   author = {Yagi, Sota and Tagami, Shunsuke},
   title = {An ancestral fold reveals the evolutionary link between {RNA} polymerase and ribosomal proteins},
   journal = {Nature Communications},
   volume = {15},
   number = {1},
   pages = {5938},
   ISSN = {2041-1723},
   DOI = {10.1038/s41467-024-50013-9},
   url = {https://doi.org/10.1038/s41467-024-50013-9},
   year = {2024},
   type = {Journal Article}
}

@article{RN24,
   author = {Zhang, Ning and Sood, Damini and Guo, Spencer C and Chen, Nanhao and Antoszewski, Adam and Marianchuk, Tegan and Dey, Supratim and Xiao, Yunxian and Hong, Lu and Peng, Xiangda},
   title = {Temperature-dependent fold-switching mechanism of the circadian clock protein {KaiB}},
   journal = {Proceedings of the National Academy of Sciences},
   volume = {121},
   number = {51},
   pages = {e2412327121},
   ISSN = {0027-8424},
   doi = {10.1073/pnas.2412327121},
   URL = {https://www.pnas.org/doi/abs/10.1073/pnas.2412327121},
   year = {2024},
   type = {Journal Article}
}

@article{RN25,
   author = {Berman, Helen M. and Burley, Stephen K.},
   title = {Protein {Data} {Bank} {(PDB)}: {Fifty}-three years young and having a transformative impact on science and society},
   journal = {Quarterly Reviews of Biophysics},
   volume = {58},
   pages = {e9},
   ISSN = {0033-5835},
   DOI = {10.1017/S0033583525000034},
   url = {https://www.cambridge.org/core/product/36E04BF6CB3DE157DF3E09EFD8EC47DC},
   year = {2025},
   type = {Journal Article}
}

@article{RN23,
   author = {Cai, Mengli and Ying, Jinfa and Lopez, Juan M and Huang, Ying and Clore, G Marius},
   title = {Unraveling structural transitions and kinetics along the fold-switching pathway of the {RfaH} {C-terminal} domain using exchange-based {NMR}},
   journal = {Proceedings of the National Academy of Sciences},
   volume = {122},
   number = {20},
   pages = {e2506441122},
   ISSN = {0027-8424},
   doi = {10.1073/pnas.2506441122},
   URL = {https://www.pnas.org/doi/abs/10.1073/pnas.2506441122},
   year = {2025},
   type = {Journal Article}
}

@article{RN49,
   author = {Chakravarty, D. and Lee, M. and Porter, L. L.},
   title = {Proteins with alternative folds reveal blind spots in {AlphaFold}-based protein structure prediction},
   journal = {Curr Opin Struct Biol},
   volume = {90},
   pages = {102973},
    ISSN = {0959-440X (Print)
0959-440x},
    doi = {https://doi.org/10.1016/j.sbi.2024.102973},
    url = {https://www.sciencedirect.com/science/article/pii/S0959440X24002008},
    year = {2025},
    type = {Journal Article}
}

@article{RN46,
   author = {Lee, M. and Schafer, J. W. and Prabakaran, J. and Chakravarty, D. and Clore, M. F. and Porter, L. L.},
   title = {Large-scale predictions of alternative protein conformations by {AlphaFold2}-based sequence association},
   journal = {Nat Commun},
   volume = {16},
   number = {1},
   pages = {5622},
   ISSN = {2041-1723},
   DOI = {10.1038/s41467-025-60759-5},
   url = {https://doi.org/10.1038/s41467-025-60759-5},
   year = {2025},
   type = {Journal Article}
}

@article{RN62,
   author = {Nicolas, William J. and Shiriaeva, Anna and Martynowycz, Michael W. and Grey, Angus C. and Ruma, Yasmeen N. and Donaldson, Paul J. and Gonen, Tamir},
   title = {Structure of the lens {MP20} mediated adhesive junction},
   journal = {Nature Communications},
   volume = {16},
   number = {1},
   pages = {2977},
   ISSN = {2041-1723},
   DOI = {10.1038/s41467-025-57903-6},
   url = {https://doi.org/10.1038/s41467-025-57903-6},
   year = {2025},
   type = {Journal Article}
}

@article{RN36,
   author = {Schafer, J. W. and Lee, M. and Chakravarty, D. and Thole, J. F. and Chen, E. A. and Porter, L. L.},
   title = {Sequence clustering confounds {AlphaFold2}},
   journal = {Nature},
   volume = {638},
   number = {8051},
   pages = {E8-e12},
   ISSN = {0028-0836},
   DOI = {10.1038/s41586-024-08267-2},
   year = {2025},
   type = {Journal Article}
}

@article{RN71,
   author = {Song, Yuxuan and Zhang, Chengxin and Omenn, Gilbert S. and O’Meara, Matthew J. and Welch, Joshua D.},
   title = {Predicting the structural impact of human alternative splicing},
   journal = {Genome Biology},
   volume = {26},
   number = {1},
   pages = {283},
   ISSN = {1474-760X},
   DOI = {10.1186/s13059-025-03744-x},
   url = {https://doi.org/10.1186/s13059-025-03744-x},
   year = {2025},
   type = {Journal Article}
}

@article{RN55,
   author = {Woodward, C. H. and Solieva, S. O. and Hwang, D. and De Paula, V. S. and Fabilane, C. S. and Young, M. C. and Trent, T. and Teeley, E. C. and Majumdar, A. and Spangler, J. B. and Bowman, G. R. and Sgourakis, N. G.},
   title = {Regulating {IL-2} Immune Signaling Function Via A Core Allosteric Structural Network},
   journal = {J Mol Biol},
   volume = {437},
   number = {2},
   pages = {168892},
   ISSN = {0022-2836 (Print)
0022-2836},
   DOI = {10.1016/j.jmb.2024.168892},
   year = {2025},
   type = {Journal Article}
}

@article{RN67,
   author = {Wu, Tianqi and Stein, Richard A. and Kao, Te-Yu and Brown, Benjamin and McHaourab, Hassane S.},
   title = {Modeling protein conformational ensembles by guiding {AlphaFold2} with {Double Electron Electron Resonance (DEER)} distance distributions},
   journal = {Nature Communications},
   volume = {16},
   number = {1},
   pages = {7107},
   ISSN = {2041-1723},
   DOI = {10.1038/s41467-025-62582-4},
   url = {https://doi.org/10.1038/s41467-025-62582-4},
   year = {2025},
   type = {Journal Article}
}

@article{RN43,
   author = {Chen, Xingyu and Stroh, Kai Steffen and Erzberger, Jan and Stengel, Florian and Pellarin, Riccardo},
   title = {Interpreting chemical crosslinks: Score-based approaches and deep neural networks},
   journal = {Current Opinion in Structural Biology},
   volume = {97},
   pages = {103237},
   ISSN = {0959-440X},
   DOI = {https://doi.org/10.1016/j.sbi.2026.103237},
   url = {https://www.sciencedirect.com/science/article/pii/S0959440X26000199},
   year = {2026},
   type = {Journal Article}
}

@article{RN66,
   author = {Feldman, Jonathan and Brogi, Maximilian and Skolnick, Jeffrey},
   title = {Adversarial Sequence Mutations in {AlphaFold} and {ESMFold} Reveal Nonphysical Structural Invariance, Confidence Failures, and Concerns for Protein Design},
   journal = {bioRxiv},
   pages = {2026.02.25.708002},
   DOI = {10.64898/2026.02.25.708002},
   url = {https://www.biorxiv.org/content/biorxiv/early/2026/02/26/2026.02.25.708002.full.pdf},
   year = {2026},
   type = {Journal Article}
}

@article{RN54,
   author = {Zhu, Kai and Trizio, Enrico and Zhang, Jintu and Hu, Renling and Jiang, Linlong and Hou, Tingjun and Bonati, Luigi},
   title = {Enhanced Sampling in the Age of Machine Learning: Algorithms and Applications},
   journal = {Chemical Reviews},
   volume = {126},
   number = {1},
   pages = {671-713},
   ISSN = {0009-2665},
   DOI = {10.1021/acs.chemrev.5c00700},
   url = {https://doi.org/10.1021/acs.chemrev.5c00700},
   year = {2026},
   type = {Journal Article}
}

@article{RN75,
    author = {He, Yao and Li, Annie Si Cong and Cai, Xiaoying and Tachiyama, Shoichi and Kumar, Rajeev and Chakravarty, Devlina and Porter, Lauren L. and Liu, Jun and Davidson, Alan R. and Zhou, Z. Hong},
    title = {A dramatic protein fold switch powers a bactericidal nanomachine},
    journal = {bioRxiv},
    DOI = {10.64898/2026.01.28.702463},
    url = {https://www.biorxiv.org/content/early/2026/01/29/2026.01.28.702463},
    year = {2026},
    type = {Journal Article}
}

@article{RN83,
  title={Dimer dissociation is a key energetic event in the fold-switch pathway of {KaiB}},
  author={Rivera, Maira and Galaz-Davison, Pablo and Retamal-Farf{\'a}n, Ignacio and Komives, Elizabeth A and Ram{\'\i}rez-Sarmiento, C{\'e}sar A},
  journal={Biophysical Journal},
  volume={121},
  number={6},
  pages={943--955},
  year={2022},
  publisher={Elsevier},
  DOI={10.1016/j.bpj.2022.02.012},
  url={https://doi.org/10.1016/j.bpj.2022.02.012}
}

@article{RN85,
  title={Extant fold-switching proteins are widespread},
  author={Porter, Lauren L and Looger, Loren L},
  journal={Proceedings of the National Academy of Sciences},
  volume={115},
  number={23},
  pages={5968--5973},
  year={2018},
  publisher={National Academy of Sciences}
}

@incollection{RN86,
  title={The many faces of structure-based potentials: from protein folding landscapes to structural characterization of complex biomolecules},
  author={Noel, Jeffrey K and Onuchic, Jos{\'e} N},
  booktitle={Computational Modeling of Biological Systems: From Molecules to Pathways},
  pages={31--54},
  year={2012},
  publisher={Springer},
  DOI={10.1073/pnas.1800168115},
  url={https://doi.org/10.1073/pnas.1800168115},
  editor={others}
}

@article{RN87,
  title={Exploring the structural acrobatics of fold-switching proteins using simplified structure-based models},
  author={Retamal-Farf{\'a}n, Ignacio and Gonz{\'a}lez-Higueras, Jorge and Galaz-Davison, Pablo and Rivera, Maira and Ram{\'\i}rez-Sarmiento, C{\'e}sar A},
  journal={Biophysical Reviews},
  volume={15},
  number={4},
  pages={787--799},
  year={2023},
  publisher={Springer},
  DOI={10.1007/s12551-023-01087-0},
  url={https://doi.org/10.1007/s12551-023-01087-0}
}

@article{
RN89,
author = {Catherine R. Knoverek  and Upasana L. Mallimadugula  and Sukrit Singh  and Enrico Rennella  and Thomas E. Frederick  and Tairan Yuwen  and Shreya Raavicharla  and Lewis E. Kay  and Gregory R. Bowman },
title = {Opening of a cryptic pocket in $\beta$-lactamase increases penicillinase activity},
journal = {Proceedings of the National Academy of Sciences},
volume = {118},
number = {47},
pages = {e2106473118},
year = {2021},
doi = {10.1073/pnas.2106473118},
URL = {https://www.pnas.org/doi/abs/10.1073/pnas.2106473118},
eprint = {https://www.pnas.org/doi/pdf/10.1073/pnas.2106473118}}

@article {RN90,
    article_type = {journal},
    title = {Opening and closing of a cryptic pocket in {VP35} toggles it between two different {RNA}-binding modes},
    author = {Mallimadugula, Upasana L and Cruz, Matthew A and Vithani, Neha and Zimmerman, Maxwell I and Bowman, Gregory R},
    editor = {Thukral, Lipi and Cui, Qiang},
    volume = 14,
    year = 2025,
    month = {sep},
    pub_date = {2025-09-02},
    pages = {RP104514},
    citation = {eLife 2025;14:RP104514},
    doi = {10.7554/eLife.104514},
    url = {https://doi.org/10.7554/eLife.104514},
    journal = {eLife},
    issn = {2050-084X},
    publisher = {eLife Sciences Publications, Ltd},}

@article{RN91,
	title = {K-{Ras}({G12C}) inhibitors allosterically control {GTP} affinity and effector interactions},
	volume = {503},
	copyright = {http://www.springer.com/tdm},
	issn = {0028-0836, 1476-4687},
	url = {https://www.nature.com/articles/nature12796},
	doi = {10.1038/nature12796},
	language = {en},
	number = {7477},
	urldate = {2026-06-09},
	journal = {Nature},
	author = {Ostrem, Jonathan M. and Peters, Ulf and Sos, Martin L. and Wells, James A. and Shokat, Kevan M.},
	month = nov,
	year = {2013},
	pages = {548--551}}

@article{RN92,
    author = {Lanman, Brian A. and Allen, Jennifer R. and Allen, John G. and Amegadzie, Albert K. and Ashton, Kate S. and Booker, Shon K. and Chen, Jian Jeffrey and Chen, Ning and Frohn, Michael J. and Goodman, Guy and Kopecky, David J. and Liu, Longbin and Lopez, Patricia and Low, Jonathan D. and Ma, Vu and Minatti, Ana E. and Nguyen, Thomas T. and Nishimura, Nobuko and Pickrell, Alexander J. and Reed, Anthony B. and Shin, Youngsook and Siegmund, Aaron C. and Tamayo, Nuria A. and Tegley, Christopher M. and Walton, Mary C. and Wang, Hui-Ling and Wurz, Ryan P. and Xue, May and Yang, Kevin C. and Achanta, Pragathi and Bartberger, Michael D. and Canon, Jude and Hollis, L. Steven and McCarter, John D. and Mohr, Christopher and Rex, Karen and Saiki, Anne Y. and San Miguel, Tisha and Volak, Laurie P. and Wang, Kevin H. and Whittington, Douglas A. and Zech, Stephan G. and Lipford, J. Russell and Cee, Victor J.},
    title = {Discovery of a Covalent Inhibitor of KRASG12C (AMG 510) for the Treatment of Solid Tumors},
    journal = {Journal of Medicinal Chemistry},
    volume = {63},
    number = {1},
    pages = {52-65},
    year = {2020},
    doi = {10.1021/acs.jmedchem.9b01180},
    note ={PMID: 31820981}, URL = {https://doi.org/10.1021/acs.jmedchem.9b01180},eprint = {https://doi.org/10.1021/acs.jmedchem.9b01180}}

@article {RN93,
article_type = {journal},
title = {Drug specificity and affinity are encoded in the probability of cryptic pocket opening in myosin motor domains},
author = {Meller, Artur and Lotthammer, Jeffrey M and Smith, Louis G and Novak, Borna and Lee, Lindsey A and Kuhn, Catherine C and Greenberg, Lina and Leinwand, Leslie A and Greenberg, Michael J and Bowman, Gregory R},
editor = {Hamelberg, Donald and Faraldo-Gómez, José D},
volume = 12,
year = 2023,
month = {jan},
pub_date = {2023-01-27},
pages = {e83602},
citation = {eLife 2023;12:e83602},
doi = {10.7554/eLife.83602},
url = {https://doi.org/10.7554/eLife.83602},
journal = {eLife},
issn = {2050-084X},
publisher = {eLife Sciences Publications, Ltd}}

@article{RN95,
  title={Conformational transitions of adenylate kinase: switching by cracking},
  author={Whitford, Paul C and Miyashita, Osamu and Levy, Yaakov and Onuchic, Jos{\'e} N},
  journal={Journal of molecular biology},
  volume={366},
  number={5},
  pages={1661--1671},
  year={2007},
  publisher={Elsevier}
}

@article{RN96,
  title={Massive conformation change in the prion protein: using dual-basin structure-based models to find misfolding pathways},
  author={Singh, Jesse P and Whitford, Paul C and Hayre, NR and Onuchic, Jos{\'e} and Cox, Daniel L},
  journal={Proteins: Structure, Function, and Bioinformatics},
  volume={80},
  number={5},
  pages={1299--1307},
  year={2012},
  publisher={Wiley Online Library},
  DOI={10.1016/j.jmb.2006.11.085},
  url={https://doi.org/10.1016/j.jmb.2006.11.085}
}

@article{RN97,
  title={Lymphotactin: how a protein can adopt two folds},
  author={Camilloni, Carlo and Sutto, Ludovico},
  journal={The Journal of chemical physics},
  volume={131},
  number={24},
  year={2009},
  publisher={AIP Publishing},
  DOI={10.1063/1.3276284},
  url={https://doi.org/10.1063/1.3276284}
}

@article{RN98,
  title={Mapping the energy landscape of a fold-switching protein {MAD2}},
  author={Pereira, Ander F and Contessoto, Vin{\'\i}cius G and Mart{\'\i}nez, Leandro and Onuchic, Jos{\'e} N},
  journal={Protein Science},
  volume={34},
  number={11},
  pages={e70329},
  year={2025},
  publisher={Wiley Online Library},
  DOI={10.1002/pro.70329},
  url={https://doi.org/10.1002/pro.70329}
}

@article{RN99,
  title={A contact-based analysis of local energetic frustration dynamics identifies key residues enabling {RfaH} fold-switch},
  author={Gonz{\'a}lez-Higueras, Jorge and Freiberger, Mar{\'\i}a In{\'e}s and Galaz-Davison, Pablo and Parra, R Gonzalo and Ram{\'\i}rez-Sarmiento, C{\'e}sar A},
  journal={Protein Science},
  volume={33},
  number={10},
  pages={e5182},
  year={2024},
  publisher={Wiley Online Library},
  DOI={10.1002/pro.5182},
  url={https://doi.org/10.1002/pro.5182}
}

@article{RN100,
  title={Molecular dynamics investigations of the $\alpha$-helix to $\beta$-barrel conformational transformation in the {RfaH} transcription factor},
  author={Gc, Jeevan B and Bhandari, Yuba R and Gerstman, Bernard S and Chapagain, Prem P},
  journal={The Journal of Physical Chemistry B},
  volume={118},
  number={19},
  pages={5101--5108},
  year={2014},
  publisher={ACS Publications},
  DOI={10.1021/jp502193v},
  url={https://doi.org/10.1021/jp502193v}
}

@article{RN101,
  title={Mechanism of the all-$\alpha$ to all-$\beta$ conformational transition of {RfaH-CTD}: Molecular dynamics simulation and markov state model},
  author={Li, Shanshan and Xiong, Bing and Xu, Yuan and Lu, Tao and Luo, Xiaomin and Luo, Cheng and Shen, Jingkang and Chen, Kaixian and Zheng, Mingyue and Jiang, Hualiang},
  journal={Journal of Chemical Theory and Computation},
  volume={10},
  number={6},
  pages={2255--2264},
  year={2014},
  publisher={ACS Publications},
  DOI={10.1021/ct5002279},
  url={https://doi.org/10.1021/ct5002279}
}

@article{RN102,
  title={Predicting protein conformational motions using energetic frustration analysis and {AlphaFold2}},
  author={Guan, Xingyue and Tang, Qian-Yuan and Ren, Weitong and Chen, Mingchen and Wang, Wei and Wolynes, Peter G and Li, Wenfei},
  journal={Proceedings of the National Academy of Sciences},
  volume={121},
  number={35},
  pages={e2410662121},
  year={2024},
  publisher={National Academy of Sciences},
  DOI={10.1073/pnas.2410662121},
  url={https://doi.org/10.1073/pnas.2410662121}
}

@article{RN103,
  title={Movie of the structural changes during a catalytic cycle of nucleoside monophosphate kinases},
  author={Vonrhein, Clemens and Schlauderer, Gerd J and Schulz, Georg E},
  journal={Structure},
  volume={3},
  number={5},
  pages={483--490},
  year={1995},
  publisher={Elsevier},
  DOI={10.1016/s0969-2126(01)00181-2},
  url={https://doi.org/10.1016/s0969-2126(01)00181-2}
}

@article{RN104,
  title={Scalable emulation of protein equilibrium ensembles with generative deep learning},
  author={Lewis, Sarah and Hempel, Tim and Jim{\'e}nez-Luna, Jos{\'e} and Gastegger, Michael and Xie, Yu and Foong, Andrew YK and Satorras, Victor Garc{\'\i}a and Abdin, Osama and Veeling, Bastiaan S and Zaporozhets, Iryna and others},
  journal={Science},
  volume={389},
  number={6761},
  pages={eadv9817},
  year={2025},
  publisher={American Association for the Advancement of Science},
  DOI={10.1126/science.adv9817},
  url={https://doi.org/10.1126/science.adv9817}
}

@article{RN105,
  title={Accelerated Sampling of Protein Dynamics Using BioEmu-Augmented Molecular Simulation},
  author={Bhakat, Soumendranath and Strauch, Eva-Maria},
  journal={Journal of Chemical Information and Modeling},
  year={2026},
  publisher={ACS Publications},
  DOI={10.1021/acs.jcim.6c01000},
  url={https://doi.org/10.1021/acs.jcim.6c01000}
}

@article{RN107,
  title = {Simple {{Allosteric Model}} for {{Membrane Pumps}}},
  author = {Jardetzky, Oleg},
  year = 1966,
  month = aug,
  journal = {Nature},
  volume = {211},
  number = {5052},
  pages = {969--970},
  publisher = {Nature Publishing Group},
  issn = {1476-4687},
  doi = {10.1038/211969a0},
  urldate = {2024-06-12},
  langid = {english}
}

@article{RN111,
  title={Atomic-level characterization of disordered protein ensembles},
  author={Mittag, Tanja and Forman-Kay, Julie D},
  journal={Current opinion in structural biology},
  volume={17},
  number={1},
  pages={3--14},
  year={2007},
  publisher={Elsevier}
}

@article{RN113,
  title={Structural plasticity as a driver of the maturation of pro-interleukin-18},
  author={Bonin, Jeffrey P and Aramini, James M and Kay, Lewis E},
  journal={Journal of the American Chemical Society},
  volume={146},
  number={44},
  pages={30281--30293},
  year={2024},
  publisher={ACS Publications}
}

@article{RN114,
  title={Making invisible excited-state structures of pro-interleukin-18 visible by combining NMR and machine learning},
  author={Bonin, Jeffrey P and Lee, Jin Sub and Liu, Zi Hao and Kim, Philip M and Kay, Lewis E},
  journal={Proceedings of the National Academy of Sciences},
  volume={123},
  number={16},
  pages={e2537014123},
  year={2026},
  publisher={National Academy of Sciences}
}

@article{RN115,
  title = {A Hierarchy of Timescales in Protein Dynamics Is Linked to Enzyme Catalysis},
  author = {{Henzler-Wildman}, Katherine A. and Lei, Ming and Thai, Vu and Kerns, S. Jordan and Karplus, Martin and Kern, Dorothee},
  year = 2007,
  month = dec,
  journal = {Nature},
  volume = {450},
  number = {7171},
  pages = {913--916},
  publisher = {Nature Publishing Group},
  issn = {1476-4687},
  doi = {10.1038/nature06407},
  urldate = {2025-02-05},
  langid = {english}
}

@article{RN116,
  title={Molecular insights into the biased signaling mechanism of the $\mu$-opioid receptor},
  author={Cong, Xiaojing and Maurel, Damien and Demene, Helene and Vasiliauskaite-Brooks, Ieva and Hagelberger, Joanna and Peysson, Fanny and Saint-Paul, Julie and Golebiowski, Jerome and Granier, Sebastien and Sounier, Remy},
  journal={Molecular Cell},
  volume={81},
  number={20},
  pages={4165--4175},
  year={2021},
  publisher={Elsevier}
}

@article{RN117,
  title={Dynamic allostery in substrate binding by human thymidylate synthase},
  author={Bonin, Jeffrey P and Sapienza, Paul J and Lee, Andrew L},
  journal={Elife},
  volume={11},
  pages={e79915},
  year={2022},
  publisher={eLife Sciences Publications, Ltd}
}

@article{RN120,
  title={Engineering a single-agent cytokine/antibody fusion that selectively expands regulatory T cells for autoimmune disease therapy},
  author={Spangler, Jamie B and Trotta, Eleonora and Tomala, Jakub and Peck, Ariana and Young, Tracy A and Savvides, Christina S and Silveria, Stephanie and Votavova, Petra and Salafsky, Joshua and Pande, Vijay S and others},
  journal={The Journal of Immunology},
  volume={201},
  number={7},
  pages={2094--2106},
  year={2018},
  publisher={Oxford University Press}
}

@article{RN122,
  title={The structural influence of the oncogenic driver mutation N642H in the STAT5B SH2 domain},
  author={Haas-Neill, Liam and Meneksedag-Erol, Deniz and Chaudhry, Ayesha and Novoselova, Masha and Ashraf, Qirat F and de Araujo, Elvin D and Wilson, Derek J and Rauscher, Sarah},
  journal={Protein Science},
  volume={34},
  number={1},
  pages={e70022},
  year={2025},
  publisher={Wiley Online Library}
}

@article{RN124,
  title={Functional protein dynamics in a crystal},
  author={Klyshko, Eugene and Kim, Justin Sung-Ho and McGough, Lauren and Valeeva, Victoria and Lee, Ethan and Ranganathan, Rama and Rauscher, Sarah},
  journal={Nature Communications},
  volume={15},
  number={1},
  pages={3244},
  year={2024},
  publisher={Nature Publishing Group UK London}
}

@article{RN125,
  title={Hidden alternative structures of proline isomerase essential for catalysis},
  author={Fraser, James S and Clarkson, Michael W and Degnan, Sheena C and Erion, Renske and Kern, Dorothee and Alber, Tom},
  journal={Nature},
  volume={462},
  number={7273},
  pages={669--673},
  year={2009},
  publisher={Nature Publishing Group UK London}
}

@article{RN127,
  title={Experimental accuracy in protein structure refinement via molecular dynamics simulations},
  author={Heo, Lim and Feig, Michael},
  journal={Proceedings of the National Academy of Sciences},
  volume={115},
  number={52},
  pages={13276--13281},
  year={2018},
  publisher={National Academy of Sciences}
}

@article{RN129,
  title={The need to implement FAIR principles in biomolecular simulations},
  author={Amaro, Rommie E and {\AA}qvist, Johan and Bahar, Ivet and Battistini, Federica and Bellaiche, Adam and Beltran, Daniel and Biggin, Philip C and Bonomi, Massimiliano and Bowman, Gregory R and Bryce, Richard A and others},
  journal={Nature methods},
  volume={22},
  number={4},
  pages={641--645},
  year={2025},
  publisher={Nature Publishing Group US New York}
}

@article{Liang2016,
  title = {Acid Activation Mechanism of the Influenza {{A M2}} Proton Channel},
  author = {Liang, Ruibin and Swanson, Jessica M.J. and Madsen, Jesper J. and Hong, Mei and De Grado, William F. and Voth, Gregory A.},
  year = 2016,
  month = nov,
  journal = {Proceedings of the National Academy of Sciences of the United States of America},
  volume = {113},
  number = {45},
  pages = {E6955-E6964},
  issn = {10916490},
  doi = {10.1073/pnas.1615471113}
}

@article{Morrison2015,
  title = {Asymmetric Protonation of {{EmrE}}},
  author = {Morrison, Emma A. and Robinson, Anne E. and Liu, Yongjia and {Henzler-Wildman}, Katherine A.},
  year = 2015,
  month = dec,
  journal = {Journal of General Physiology},
  volume = {146},
  number = {6},
  pages = {445--461},
  issn = {1540-7748},
  doi = {10.1085/jgp.201511404}
}

@article{Smith1998,
  title = {A Novel Calcium-Sensitive Switch Revealed by the Structure of Human {{S100B}} in the Calcium-Bound Form},
  author = {Smith, Steven P and Shaw, Gary S},
  year = 1998,
  month = feb,
  journal = {Structure},
  volume = {6},
  number = {2},
  pages = {211--222},
  issn = {0969-2126},
  doi = {10.1016/S0969-2126(98)00022-7},
  urldate = {2026-06-22}
}

@article{Swanson2007,
  title = {Proton {{Solvation}} and {{Transport}} in {{Aqueous}} and {{Biomolecular Systems}}:\, {{Insights}} from {{Computer Simulations}}},
  shorttitle = {Proton {{Solvation}} and {{Transport}} in {{Aqueous}} and {{Biomolecular Systems}}},
  author = {Swanson, Jessica M. J. and Maupin, C. Mark and Chen, Hanning and Petersen, Matt K. and Xu, Jiancong and Wu, Yujie and Voth, Gregory A.},
  year = 2007,
  month = may,
  journal = {The Journal of Physical Chemistry B},
  volume = {111},
  number = {17},
  pages = {4300--4314},
  publisher = {American Chemical Society},
  issn = {1520-6106},
  doi = {10.1021/jp070104x},
  urldate = {2024-06-13}
}

@article{Trave1995,
  title = {Molecular Mechanism of the Calcium-Induced Conformational Change in the Spectrin {{EF-hands}}},
  author = {Trav{\'e}, G. and Lacombe, P. J. and Pfuhl, M. and Saraste, M. and Pastore, A.},
  year = 1995,
  month = oct,
  journal = {The EMBO journal},
  volume = {14},
  number = {20},
  pages = {4922--4931},
  issn = {0261-4189},
  doi = {10.1002/j.1460-2075.1995.tb00175.x},
  langid = {english},
  pmcid = {PMC394594},
  pmid = {7588621}
}

@article{Wright2005,
  title = {The {{Three-dimensional Solution Structure}} of {{Ca2}}+-Bound {{S100A1}} as {{Determined}} by {{NMR Spectroscopy}}},
  author = {Wright, Nathan T. and Varney, Kristen M. and Ellis, Karen C. and Markowitz, Joseph and Gitti, Rossitza K. and Zimmer, Danna B. and Weber, David J.},
  year = 2005,
  month = oct,
  journal = {Journal of Molecular Biology},
  volume = {353},
  number = {2},
  pages = {410--426},
  issn = {0022-2836},
  doi = {10.1016/j.jmb.2005.08.027},
  urldate = {2026-06-22}
}

@article{Teng2025,
  title = {{AF2RAVE}: Protein Ensemble Generation with Physics-Based Sampling},
  shorttitle = {Af2rave},
  author = {Teng, Da and Meraz, Vanessa J. and Aranganathan, Akashnathan and Gu, Xinyu and Tiwary, Pratyush},
  year = 2025,
  month = jul,
  journal = {Digital Discovery},
  volume = {4},
  pages = {2052--2061},
  publisher = {RSC},
  issn = {2635-098X},
  doi = {10.1039/D5DD00201J},
  urldate = {2025-07-15},
  langid = {english}
}

@article{Aranganathan2026,
  title = {Applied {{Causality}} to {{Infer Protein Dynamics}} and {{Kinetics}}},
  author = {Aranganathan, Akashnathan and Beyerle, Eric R.},
  year = 2026,
  month = feb,
  journal = {Journal of Chemical Information and Modeling},
  volume = {66},
  number = {3},
  pages = {1661--1674},
  publisher = {American Chemical Society},
  issn = {1549-9596},
  doi = {10.1021/acs.jcim.5c02554},
  urldate = {2026-06-23}
}

@article{Zheng2024,
  title = {Predicting Equilibrium Distributions for Molecular Systems with Deep Learning},
  author = {Zheng, Shuxin and He, Jiyan and Liu, Chang and Shi, Yu and Lu, Ziheng and Feng, Weitao and Ju, Fusong and Wang, Jiaxi and Zhu, Jianwei and Min, Yaosen and Zhang, He and Tang, Shidi and Hao, Hongxia and Jin, Peiran and Chen, Chi and No{\'e}, Frank and Liu, Haiguang and Liu, Tie-Yan},
  year = 2024,
  month = may,
  journal = {Nature Machine Intelligence},
  volume = {6},
  number = {5},
  pages = {558--567},
  issn = {2522-5839},
  doi = {10.1038/s42256-024-00837-3},
  urldate = {2024-08-26},
  langid = {english}
}

@article{Abramson2024,
  title = {Accurate Structure Prediction of Biomolecular Interactions with {{AlphaFold}} 3},
  author = {Abramson, Josh and Adler, Jonas and Dunger, Jack and Evans, Richard and Green, Tim and Pritzel, Alexander and Ronneberger, Olaf and Willmore, Lindsay and Ballard, Andrew J. and Bambrick, Joshua and Bodenstein, Sebastian W. and Evans, David A. and Hung, Chia-Chun and O'Neill, Michael and Reiman, David and Tunyasuvunakool, Kathryn and Wu, Zachary and {\v Z}emgulyt{\.e}, Akvil{\.e} and Arvaniti, Eirini and Beattie, Charles and Bertolli, Ottavia and Bridgland, Alex and Cherepanov, Alexey and Congreve, Miles and {Cowen-Rivers}, Alexander I. and Cowie, Andrew and Figurnov, Michael and Fuchs, Fabian B. and Gladman, Hannah and Jain, Rishub and Khan, Yousuf A. and Low, Caroline M. R. and Perlin, Kuba and Potapenko, Anna and Savy, Pascal and Singh, Sukhdeep and Stecula, Adrian and Thillaisundaram, Ashok and Tong, Catherine and Yakneen, Sergei and Zhong, Ellen D. and Zielinski, Michal and {\v Z}{\'i}dek, Augustin and Bapst, Victor and Kohli, Pushmeet and Jaderberg, Max and Hassabis, Demis and Jumper, John M.},
  year = 2024,
  month = jun,
  journal = {Nature},
  volume = {630},
  number = {8016},
  pages = {493--500},
  publisher = {Nature Publishing Group},
  issn = {1476-4687},
  doi = {10.1038/s41586-024-07487-w},
  urldate = {2024-12-11},
  langid = {english}
}

@article{Qiao2024,
  title = {State-Specific Protein--Ligand Complex Structure Prediction with a Multiscale Deep Generative Model},
  author = {Qiao, Zhuoran and Nie, Weili and Vahdat, Arash and Miller, Thomas F. and Anandkumar, Animashree},
  year = 2024,
  month = feb,
  journal = {Nature Machine Intelligence},
  volume = {6},
  number = {2},
  pages = {195--208},
  publisher = {Nature Publishing Group},
  issn = {2522-5839},
  doi = {10.1038/s42256-024-00792-z},
  urldate = {2024-08-07},
  langid = {english}
}

@article{Krishna2024,
  title = {Generalized Biomolecular Modeling and Design with {{RoseTTAFold All-Atom}}},
  author = {Krishna, Rohith and Wang, Jue and Ahern, Woody and Sturmfels, Pascal and Venkatesh, Preetham and Kalvet, Indrek and Lee, Gyu Rie and {Morey-Burrows}, Felix S. and Anishchenko, Ivan and Humphreys, Ian R. and McHugh, Ryan and Vafeados, Dionne and Li, Xinting and Sutherland, George A. and Hitchcock, Andrew and Hunter, C. Neil and Kang, Alex and Brackenbrough, Evans and Bera, Asim K. and Baek, Minkyung and DiMaio, Frank and Baker, David},
  year = 2024,
  month = mar,
  journal = {Science},
  volume = {384},
  number = {6693},
  pages = {eadl2528},
  publisher = {American Association for the Advancement of Science},
  doi = {10.1126/science.adl2528},
  urldate = {2025-01-29}
}

@article{Masters2025,
  title = {Investigating Whether Deep Learning Models for Co-Folding Learn the Physics of Protein-Ligand Interactions},
  author = {Masters, Matthew R. and Mahmoud, Amr H. and Lill, Markus A.},
  year = 2025,
  month = oct,
  journal = {Nature Communications},
  volume = {16},
  number = {1},
  pages = {8854},
  publisher = {Nature Publishing Group},
  issn = {2041-1723},
  doi = {10.1038/s41467-025-63947-5},
  urldate = {2025-10-30},
  langid = {english}
}

@article{thakur2024potts,
  title={Potts {Hamiltonian} models and molecular dynamics free energy simulations for predicting the impact of mutations on protein kinase stability},
  author={Thakur, Abhishek and Gizzio, Joan and Levy, Ronald M},
  journal={The Journal of Physical Chemistry B},
  volume={128},
  number={7},
  pages={1656--1667},
  year={2024},
  publisher={ACS Publications}
}

@article{vardanyan2026results,
  title={Results of a Large-Scale Study of the Binding of 50 Type {II} Inhibitors to 348 Kinases: The Role of Protein Reorganization},
  author={Vardanyan, Vardan H and Haldane, Allan and Hwang, Howook and Coskun, Dilek and Lihan, Muyun and Miller, Edward B and Friesner, Richard A and Levy, Ronald M},
  journal={Journal of Medicinal Chemistry},
  volume={69},
  number={8},
  pages={9507--9520},
  year={2026},
  publisher={ACS Publications}
}

@article{stein2022speach_af,
  title={SPEACH\_AF: Sampling protein ensembles and conformational heterogeneity with Alphafold2},
  author={Stein, Richard A and Mchaourab, Hassane S},
  journal={PLoS computational biology},
  volume={18},
  number={8},
  pages={e1010483},
  year={2022},
  publisher={Public Library of Science San Francisco, CA USA}
}

@article{kalakoti2025afsample2,
  title={AFsample2 predicts multiple conformations and ensembles with AlphaFold2},
  author={Kalakoti, Yogesh and Wallner, Bj{\"o}rn},
  journal={Communications biology},
  volume={8},
  number={1},
  pages={373},
  year={2025},
  publisher={Nature Publishing Group UK London}
}

@article{volk2023alphaflow,
  title={AlphaFlow: autonomous discovery and optimization of multi-step chemistry using a self-driven fluidic lab guided by reinforcement learning},
  author={Volk, Amanda A and Epps, Robert W and Yonemoto, Daniel T and Masters, Benjamin S and Castellano, Felix N and Reyes, Kristofer G and Abolhasani, Milad},
  journal={Nature Communications},
  volume={14},
  number={1},
  pages={1403},
  year={2023},
  publisher={Nature Publishing Group UK London}
}

@article {Clore2025.12.30.697132,
        author = {Clore, Madeleine F. and Thole, Joseph F. and Dontha, Suchetan and Sharma, Pramesh and Greenberg, Naomi and Strub, Marie-Paule and Starich, Mary and Ott, Carolyn and Jensen, Davin and Volkman, Brian F. and Coudron, Matthew and Porter, Lauren L.},
        title = {AlphaFold reveals but sometimes distorts an organizational principle of protein folding},
        elocation-id = {2025.12.30.697132},
        year = {2026},
        doi = {10.64898/2025.12.30.697132},
        publisher = {Cold Spring Harbor Laboratory},
        URL = {https://www.biorxiv.org/content/early/2026/07/21/2025.12.30.697132},
        eprint = {https://www.biorxiv.org/content/early/2026/07/21/2025.12.30.697132.full.pdf},
        journal = {bioRxiv}
}

@article{kim2025large,
  title={Large scale prospective evaluation of co-folding across 557 Mac1-ligand complexes and three virtual screens},
  author={Kim, Jongbin and Correy, Galen J and Hall, Brendan W and Rachman, Moira M and Mailhot, Olivier and Togo, Takaya and Gonciarz, Ryan L and Jaishankar, Priyadarshini and Neitz, R Jeffrey and Hantz, Eric R and others},
  journal={bioRxiv},
  pages={2025--12},
  year={2025},
  publisher={Cold Spring Harbor Laboratory}
}

@article{yu2026bias,
  title={Bias in the AlphaFold3 prediction of ligand-induced domain motion in enzymes},
  author={Yu, Hao and Bekar-Cesaretli, Ayse A and Lazou, Maria and Kozakov, Dima and Joseph-McCarthy, Diane and Vajda, Sandor},
  journal={Proceedings of the National Academy of Sciences},
  volume={123},
  number={10},
  pages={e2530709123},
  year={2026},
  publisher={National Academy of Sciences}
}

@article{anthis2015visualizing,
  title={Visualizing transient dark states by NMR spectroscopy},
  author={Anthis, Nicholas J and Clore, G Marius},
  journal={Quarterly Reviews of Biophysics},
  volume={48},
  number={1},
  pages={35--116},
  year={2015},
  publisher={Cambridge University Press}
}

@article{shan2011does,
  title={How does a drug molecule find its target binding site?},
  author={Shan, Yibing and Kim, Eric T and Eastwood, Michael P and Dror, Ron O and Seeliger, Markus A and Shaw, David E},
  journal={Journal of the American Chemical Society},
  volume={133},
  number={24},
  pages={9181--9183},
  year={2011},
  publisher={ACS Publications}
}

\end{document}